\documentclass[manuscript]{aastex}
\usepackage{rotating,booktabs}

\slugcomment{For submission to the Astrophysical Journal}

\shorttitle{Multiple Radial Cool Molecular Filaments in NGC~1275}
\shortauthors{Ho et al.}

\begin{document}


\title{Multiple Radial Cool Molecular Filaments in NGC~1275 }


\author{I-Ting Ho}
\affil{National Taiwan University \\
and \\ 
Institute of Astronomy \& Astrophysics, Academia Sinica, PO Box 23-141, Taipei 10617, Taiwan}
\email{itho@asiaa.sinica.edu.tw}

\author{Jeremy Lim}
\affil{Institute of Astronomy \& Astrophysics, Academia Sinica, PO Box 23-141, Taipei 10617, Taiwan}
\email{jlim@asiaa.sinica.edu.tw}

\and

\author{Dinh-V-Trung}
\affil{Institute of Astronomy \& Astrophysics, Academia Sinica, PO Box 23-141, Taipei 10617, Taiwan\altaffilmark{1}}
\email{trung@asiaa.sinica.edu.tw}

\altaffiltext{1}{on leave from Center for Quantum Electronics, Institute of Physics, Vietnamese Academy of Science and Technology, 10 DaoTan, BaDinh, Hanoi, Vietnam}



\begin{abstract}

We have extended our previous observation \citep{lim08} of NGC~1275 (Perseus~A, the central giant elliptical galaxy in the Perseus Cluster) covering a central radius of $\sim$10~kpc to the entire main body of cool molecular gas spanning $\sim$14~kpc east and west of center.  We find no new features beyond the region previously mapped, and show that all six spatially-resolved features on both the eastern and western sides (three on each side) comprise radially aligned filaments.  Such radial filaments can be most naturally explained by a model in which gas deposited "upstream" in localized regions experiencing a X-ray cooling flow subsequently free falls along the gravitational potential of Per~A, as we previously showed can explain the observed kinematics of the two longest filaments.  All the detected filaments coincide with locally bright H$\alpha$ features, and have a ratio in CO(2--1) to H$\alpha$ luminosity of $\sim$$10^{-3}$; we show that these filaments have lower star formation efficiencies than the nearly constant value found for molecular gas in nearby normal spiral galaxies.  On the other hand, some at least equally luminous H$\alpha$ features, including a previously identified giant HII region, show no detectable cool molecular gas with a corresponding ratio at least a factor of $\sim$5 lower; in the giant HII region, essentially all the pre-existing molecular gas may have been converted to stars.  We demonstrate that all the cool molecular filaments are gravitationally bound, and without any means of support beyond thermal pressure should collapse on timescales $\lesssim 10^6$~yrs.  By comparison, as we showed previously the two longest filaments have much longer dynamical ages of $\sim$$10^7$~yrs.  Tidal shear may help delay their collapse, but more likely turbulent velocities of at least a few tens ${\rm km \ s^{-1}}$ or magnetic fields with strengths of at least several $\sim$$10~\mu$G are required to support these filaments.

\end{abstract}


\section{INTRODUCTION}
NGC~1275 (henceforth Perseus~A, abbreviated Per~A) is the central dominant cD (giant elliptical) galaxy in the Perseus Cluster.  It is one of the most intriguing and intensively scrutinized galaxies in the sky.  \citet{min57} found two sets of optical emission lines separated by $\sim$$3000 {\rm \ km \ s^{-1}}$ towards Per~A, and proposed that Per~A is colliding with a forergound infalling galaxy.  A photograph in the H$\alpha$ line taken by \citet{lyn70} revealed a spectacular flamentary nebula centered on Per~A and extending up to $\sim$140\arcsec\ ($\sim$50~kpc) from its center.  A modern CCD image reveals that the H$\alpha$ nebula has a bright inner region extending east-west, together with a multitude of mostly radial (but also some transverse) filaments that extends furthest in the north-south direction reaching $\sim$70~kpc north of center \citep{con01}.  Per~A was the first central cD galaxy in a rich cluster found to contain cool ($\lesssim$100~K) molecular gas as traced in CO \citep{laz89,mir89}.  Single-dish maps show that the main body of the cool molecular gas has an overall morphology similar to the bright inner region of the H$\alpha$ nebula \citep{sal06}.  Cool molecular gas also has been detected towards the outer H$\alpha$ filaments where observed \citep{sal06, sal08b}.  Both hot ($\sim$2000~K) \citep[][and references therein]{hat05} and warm ($\sim$300--400~K) \citep{joh07} molecular hydrogen gas has been found towards all regions of the H$\alpha$ nebula so far observed (bright inner region and outer filaments).  The coolest detectable component of the X-ray emitting gas in the Perseus Cluster, at a temperature of $\sim$$6 \times 10^6$~K, has a morphology similar to the H$\alpha$ nebula \citep{fab06}.

The nature (origin and excitation) of the multi-phase nebula associated with Per~A is poorly understood.  Because the cool molecular gas traced in CO dominates by far the mass of this nebula, a detailed understanding of its properties is of particular importance.  The main body of this cool molecular gas has been traced by single-dish telescopes out to $\sim$14~kpc east and west of center, and has a mass of $\sim$$10^{10} {\rm \ M_\sun}$ \citep[][and references therein]{sal06}.  These single-dish observations at spatial resolutions of several kiloparsecs show no overall pattern in the gas kinematics, which is slightly redshifted at the nucleus and (mostly) more strongly blueshifted east and west of center.  Our \citep[][hereafter Paper~I]{lim08} interferometric observations at a spatial resolution of $\sim$1~kpc covering the central region of radius $\sim$10~kpc revealed three radial filaments lying approximately east-west, together with a number of other mostly unresolved features.  As a whole, these features show no overall pattern in their kinematics; in particular, there is no bulk velocity gradient in the east-west (or any other) direction indicative of gas orbiting Per~A, as would be expected for material captured from a gas rich-galaxy.  Instead, we found that each filament exhibits its own particular kinematic pattern.  The outermost filaments east and west of center both exhibit linearly increasing blueshifted velocities with decreasing radii, but with different velocity gradients.  We were able to reproduce their observed spatial-kinematic structures as free-fall in the gravitational potential of Per~A, as would be naturally expected for cool gas deposited by a X-ray cooling flow.

Here, we report follow-up interferometric CO observations that extend our coverage in the east-west direction to span the entire main body of the cool molecular gas mapped with single-dish telescopes.  We describe our observations and data reduction in $\S2$.  The results are presented in $\S3$, where we show that all but one of the detected CO features is spatially resolved into radial filaments; at least six distinct filaments can be clearly identified in our maps.  In $\S4$, we examine the relationship between the cool molecular filaments and the H$\alpha$ nebula, and show that there is a correlation between the CO and H$\alpha$ luminosity of all these filaments; on the other hand, we show that a number of at least equally luminous H$\alpha$ features exhibit no detectable CO with upper limits that deviate strongly from the derived correlation.  We then demonstrate that all the filaments are gravitationally bound, and explore why they do not appear to be collapsing to vigorously form stars.  Finally, as a guide for future interferometric observations, we identify regions where our interferometric measurements have not recovered all the single-dish emission.  In $\S5$, we provide a concise summary of our observations and interpretation.  We assume as in Paper~I that Per~A lies at a distance of $\sim$74~Mpc (redshift $z=0.01756$, and adopting $H_0 = 70 {\rm \ km \ s^{-1} \ Mpc^{-1}}$ and $\Omega=1$), so that $1\arcsec = 360 {\rm \ pc}$.

\section{OBSERVATIONS AND DATA REDUCTION}\label{observations and data reduction}
We observed Per~A on 2006 December 17 using the Sub-Millimeter Array (SMA) as in our previous observation (Paper~I).  The weather was good, with a zenith sky opacity at 225~GHz ($\tau_{\rm 225~GHz}$) in the range $\sim$0.05--0.1 as measured at the adjacent Caltech Submillimeter Observatory.  The SMA was in its compact configuration, and with six of its eight 6-m antennas operating provided fifteen baselines ranging from about 10~m to 70~m. The dual-sideband receivers operating in the 230~GHz (1.3~mm) band were tuned to a central observing frequency of 216.35~GHz in the lower sideband, and 226.34~GHz in the upper sideband to place the CO(2--1) line in this sideband.  At the frequency of the CO(2--1) line, the primary beam of the telescope as measured at full width half maximum (FWHM) is $\sim$55\arcsec, which corresponds to $\sim$20~kpc in the assumed Cosmology.  The correlator was configured to cover 24 spectral windows (chunks in the SMA nomenclature) with 128 channels in each chunk.  This provided a spectral resolution of 0.81~MHz ($\sim$$1.1 {\rm \ km \ s^{-1}}$) over a total bandwidth, with some overlap at the edges of adjacent chucks, of 1.98~GHz ($\sim$$2590 {\rm \ km \ s^{-1}}$).  The system temperature ranged from 80~K to 170~K, changing in time with the source elevation.

We began by observing Neptune and Uranus for 20 minutes each to serve as alternate choices for absolute flux and antenna passband calibration.  We then observed Per~A over a duration of $\sim$9~hrs, interrupted only once for an antenna pointing check.  Our previous observation (Paper~I) had shown that the CO(2-1) features detected spanned from center to (near) the edge of the primary beam in the east-west directions.  To provide better sensitivity to the outer eastern and western features, and to search for more gas beyond the field of view of our previous observation, we employed a three-point mosaic with one field centered on the nucleus and the other two centered approximately on the Eastern and Western filaments respectively as illustrated in Figure~\ref{3pointing}.  We pointed the telescope alternately between these three positions, and in each cycle integrated for six 30-s samples on the central position and ten 30-s samples on each of the eastern and western positions.  More details of the observations can be found in Table~\ref{pointing}.  Finally, we observed Saturn for 55~min to serve as another alternative for passband calibration.  Saturn was so bright, however, that it partially saturated the receivers, thus preventing its use as a passband calibrator.

We performed our data reduction using SMA-specific MIR tasks in IDL adopted for the SMA from the MMA software package developed originally for OVRO \citep{sco93}.  Because absorption by the Earth's atmosphere introduces an elevation-dependent attenuation of astronomical signals, we first corrected for this effect by weighting the amplitude of each visibility by the inverse product of the system temperatures of the corresponding antenna pair as measured in each 30-s integration.  In our observation, both Uranus and Neptune were resolved on the longer baselines; because the measured amplitude of Uranus was higher than that of Neptune on all baselines, we used Uranus for both absolute flux and passband calibration.  We derived complex gain (i.e., amplitude and phase) solutions from the unresolved nuclear continuum emission of Per~A when the telescope was pointed at the central position, and then interpolated these solutions to measurements when the telescope was pointed at the eastern and western positions.  Because of significant baseline-based errors in some chunks, we adopted baseline-based solutions for both passband and gain calibration.  This also helped address significant pointing errors on some of the antennas, an issue we will return to later.

Mapping of the CO(2--1) emission towards Per~A requires stable antenna passbands to avoid artifacts that would otherwise be produced by its strong nuclear continuum source (see discussion in Paper~I).  This source had a measured flux density of 3.5~Jy during our observation, compared with 3.4~Jy measured in our previous observation (Paper~I).  In channel maps of the central pointing made before continuum subtraction, we found that the nuclear continuum source exhibits a slow and weak variation in intensity with frequency, reflecting small changes in the individual antenna passbands with time.  The peak-to-peak variation in the nuclear continuum intensity with frequency is $\sim$$3\%$.  This led to an imperfect subtraction of the nuclear continuum source in our CO(2--1) channel maps, leaving artifacts as strong as about $\pm$50~mJy at the nuclear position for the central pointing.  For comparison, the root-mean-square (rms) noise level ($\sigma$) of the channel maps at  a velocity resolution of $\sim$$20 {\rm \ km \ s^{-1}}$ is $\sim$30~mJy for the central pointing (Table~\ref{pointing}).  Hence, artifacts at the nuclear position are no stronger than about $\pm$2$\sigma$ in these maps.  In channel maps of the eastern and western pointings where the nuclear continuum source lies close to the edge of the primary beam and is therefore strongly attenuated, artifacts at the nuclear position caused by passband variations is correspondingly less severe.

To make the final CLEAN maps, the DIRTY BEAM, corresponding to the point spread function (PSF) of the telescope as computed from the known positions of the antennas relative to the source, must be deconvolved from the DIRTY maps (where each pixel is convolved with the telescope PSF) to remove the sidelobe response of the telescope.  Variations in antenna passbands, along with imperfect complex gain calibration, can leave artifacts related to imperfect deconvolution of sidelobes from both continuum and line emission.  Although difficult to quantify, artifacts produced by sidelobes can often be recognized by their resemblance to the known sidelobe pattern.  In addition, maps made in our previous observation (Paper~1) could be compared with those made here as an independent check.  In this way, we were able to identify artifacts lying north-south in our previous map, and confirm as genuine all the features lying east-west.

For the eastern and western pointings, the nuclear continuum source of Per~A lies close to the edge of the primary beam.  Any pointing errors in the individual antennas (along the radial direction) therefore causes a much larger variation in the measured intensity of the nuclear continuum source for these pointings compared with the central pointing.  Indeed, we found that the measured intensity of the nuclear continuum source changed with time by up to a factor of $\sim$1.3 larger in both the eastern and western pointings than in the central pointing, indicating significant antenna pointing errors.  By comparing the amplitudes of the nuclear continuum source measured in each of the three pointings against their expected amplitudes, we estimated pointing errors of about 5\arcsec--15\arcsec\ depending on the given antenna.  Such antenna pointing errors lead to an inaccurate correction for the primary beam response of the antennas that becomes increasingly worse for features closer to the edge of the primary beam.  To estimate the magnitude of these errors, we compared the measured intensity of the nuclear continuum source from maps (corrected for the primary beam response) made of adjacent pointings; we also compared, after continuum subtraction, the measured integrated intensity of CO(2--1) features in regions of the map (corrected for the primary beam response) where adjacent pointings overlap.  We found that the intensity of both the continuum source and common CO(2--1) features to differ by less than $\sim$$10\%$ between adjacent pointings, indicating that the antenna pointing errors are more or less random in the sky.  In the CO(2-1) channel maps at a velocity resolution of $\sim$$20 {\rm \ km \ s^{-1}}$, 10$\%$ variations in the intensity of individual features correspond to less than the rms noise fluctuations.

Antenna pointing errors, just like errors in passband and complex gain calibration, also can leave artifacts related to imperfect deconvolution of sidelobes from both continuum and line emission.  In this respect, antenna pointing errors are probably more important than those related to variations in antenna passbands and imperfect complex gain calibration for, especially, the eastern and western pointings.

To make CO(2--1) channel maps, we first subtracted the nuclear continuum source in the visibility plane derived from a fit to the line-free channels in the upper sideband.  For the reader to judge the reliability of our maps, we have made separate channel maps for the three individual pointings at a velocity resolution of $20 {\rm \ km \ s^{-1}}$, and derived from these channel maps separate integrated intensity (zeroth moment) and intensity-weighted mean-velocity (first moment) maps corrected for the primary beam response.  We also made a mosaic map utilizing all the data from the three separate pointings, along with data taken in the previous observation (Paper~I), with the appropriate correction for the primary beam response in each pointing.  The corresponding angular resolutions and rms noise of the mosaic and combined maps are listed in Table~\ref{pointing}.
All the maps shown in this manuscript, unless otherwise mentioned, were made with natural weighting of the data to maximize the signal-to-noise (S/N) ratio.  Only one of the maps shown was made using uniform weighting, useful for inspecting the detected features at the highest attainable angular resolution.

We have discovered a small error in our subtraction of the nuclear continuum source in Paper~I.  When deriving its flux density, we inadvertently included line-containing along with line-free channels in our fit: this resulted in a slight overestimate of the flux density of the nuclear continuum source, and therefore an oversubtraction of the continuum source in our channel maps.  This error can be seen most clearly in the spectra of the individual filaments shown in Figure~3 of Paper~I, especially for the Inner and Western filaments (i.e., those closest to the nucleus) where the spectral baseline is consistently below zero by a few tens of mJy.  After correcting this mistake, the moment map made, shown in Figure~\ref{3pointing}, does not show any qualitative changes.

As in Paper~I, to make moment maps we first convolved the channel maps with a triangular function (i.e., Hanning smoothing) that best recovers the overall features in that particular map.  We then discarded individual pixels in the original channel maps with intensities less than a certain theshold (typically $\sim$1.5--2.0$\sigma$) in their corresponding Hanning-smoothed channel maps.  We summed the remaining data in the channel maps (with no smoothing) to produce the integrated intensity map, and computed the mean velocity of each pixel weighted by its intensity in each channel to produce the intensity-weighted mean-velocity map.

\section{RESULTS}

\subsection{Individual Pointings}\label{individual pointings}
Figure~\ref{new_mom} shows moment maps of the central, eastern, and western pointings separately.  In each case, the integrated CO(2--1) intensity, corrected for the primary beam response, is plotted in contours on the intensity-weighted CO(2--1) mean velocity in color.  All of the features seen in our previous observation (comprising a single pointing centered on the nucleus) except for N1, S1, and W1 (Fig.~2 in Paper~I; see also Fig.~1 of this paper) are seen in at least one (typically two, if not all three) of the individual pointings in our present observation.  S1, if real, should have been detectable in the moment maps of both the central and western pointings.  On the other hand, N1 and W1, if real, would lie below the cutoff threshold imposed on the moment maps for both the central and western pointings.  When we lowered the cutoff threshold for making the moment maps shown in Figure~\ref{new_mom}, we found that W1 but not N1 nor S1 could be seen in both the central (along with E1) and western pointings.  Thus, W1 is real, but as shown explicitly in $\S\ref{artifacts}$ both N1 and S1 are artifacts related to sidelobes.

The Inner filament is detected in all three pointings.  The Western filament is detected in both the central and, at a higher S/N ratio, western pointings.  The Eastern filament is detected in the eastern pointing, and only its outermost tip in the central pointing.  E1 is detected only in the eastern pointing, but E2 is detected in both the central and eastern pointings.  Apart from these features, all of which were previously detected in Paper~I, we do not detect any new features in the maps of Figure~\ref{new_mom} that we can confidently regard as real.

\subsection{Mosaic of All Available Data}
We have combined all the data taken in our present observation with that taken in Paper~I (comprising only a single pointing centered on the nucleus) to make the most sensitive mosaic map possible.  Figure~\ref{channel-maps} shows the resulting CO(2--1) channel maps for the inner$+$western regions (upper panel) and the eastern region (lower panel) separately.  Figure~\ref{mom-all} shows the corresponding moment map, where the dotted line indicates where the noise level has increased by a factor of 2 from center.  Although extending the region mapped from a central radius of $\sim$10~kpc to a radial extent of $\sim$14~kpc in both the eastern and western directions to cover the entire main body of molecular gas in Per~A, we did not detect any new features beyond the region previously mapped in Paper~I.  In the moment map of Figure~\ref{mom-all}, we have masked out the artifact labeled N1 in our previous observation (see $\S\ref{artifacts}$); the artifact S1, which we found to be substantially reduced in the maps made from all the data combined, is not detected at the cutoff threshold imposed.  (We note, for completeness, that noise peaks outside the area occupied by the dotted line have been masked.) Compared with our previous observation, the Western filament appears to exhibit a more pronounced bend towards the south about halfway along its length.  As shown in greater detail later in $\S\ref{position-velocity diagrams}$, this apparent bend is caused by a separate feature labeled W2 (not discernible as a separate feature in Paper~I despite the improved continuum subtraction; Fig.~\ref{3pointing}) detected here for the first time (at the improved sensitivity) located just to the south of the Western filament about halfway along its length.

In Figure~\ref{uniform-weighting}, we show moment maps of the features E1, E2, and W1 made with uniform weighting to provide a higher angular resolution.  The arrows in each panel point to the center of Per~A.  All these features, like the three previously identified filaments, can now be seen to comprise radial filaments (as discussed in more detail below).  \citet{sal08a} have observed the Eastern filament, E1, and E2 at significantly higher sensitivity and comparable (marginally higher) angular resolution with the IRAM~PdBI.  Their maps also show that E1 and E2 are radial filaments.

In Table~\ref{gas_mass}, we tabulate the properties of each filament (with limited information for W2, which is spatially unresolved), including its angular distance from center (measured at the midpoint), length, position angle, and velocity range.  The velocity range spanned by each filament is defined as the extreme velocities at which the filaments are detectable at or above the $3\sigma$ level in the channel maps as cross-checked against the moment maps.  As mentioned in Paper~I, the lengths of the Inner, Western, Eastern, and now also E2 filaments are difficult to quantify precisely because they do not have smoothly varying intensity profiles; here, we use the same method as in Paper~I, with their lengths defined as the difference in radial positions of their emission centroids either at or close to their opposite ends in velocity (Eastern and Western filaments) or where their angular separation is largest (Inner and E2 filaments).  This method likely underestimates slightly their true lengths.  Their position angles correspond to that measured from the center of Per~A through their midpoints as estimated by eye in the moment maps.  

We note that, compared to the values previously tabulated (Table~1 in Paper~I), the largest difference is for the Inner filament which is now much longer and spans a larger velocity range.  This is because the error in continuum subtraction (as described towards the end of $\S$\ref{observations and data reduction}) most strongly affects the Inner filament, which now extends much further beyond the nucleus on the eastern side.  The Western filament is detectable over a slightly larger velocity range because the improved continuum subtraction and increased sensitivity permits previously undetectable (i.e., below $3\sigma$ in the channel maps) portion of the filament to now be detectable.  The position-velocity (PV-) diagrams of the Western and Eastern filaments have not changed qualitatively (see $\S$\ref{position-velocity diagrams}), and provide if anything a more confident fit (given the improved sensitivity) to the model presented in Paper~I whereby these filaments are in free-fall along the gravitational potential of Perseus~A.  The revised dynamical age of the Western filament is $\sim$23~Myr and Eastern filament $\sim$28~Myr; i.e., of order $\sim$$10^7$~yr as estimated in Paper~I.

The E2 and W1 filaments are best resolved in the uniformly-weighted moment maps of Figure~\ref{uniform-weighting}.  We have therefore derived their lengths and position angles by fitting gaussian structures to these filaments in the uniformly-weighted maps but without imposing any cutoffs to preserve error statistics.  We find both to be aligned radially with respect to the center of Per~A within measurement uncertainties (no more than $\sim$$1.5 \sigma$).  For W2, we tabulate only its radial distance from and its position angle relative to the center.

\subsubsection{Sidelobe Artifacts}\label{artifacts}
As mentioned in Paper~I, the CO(2--1) features detected along the east-west direction (all now confirmed to be genuine) appear to be associated with H$\alpha$ emission, but not those detected either to the north (N1, which is too weak to be detected in the present observation) or south (S1, which is not confirmed in the present observation).  To properly understand whether all the detected CO(2--1) features have H$\alpha$ counterparts, we have carefully examined whether N1 and S1 are artifacts related to imperfect removal of sidelobes.

In Figure~\ref{artifact}, we show both DIRTY (before deconvolving the telescope PSF) and CLEAN (after deconvolving the telescope PSF) maps of just the channel where S1 appears most strongly in the previous dataset.  We also show in this figure the DIRTY beam (the telescope PSF).  Apart from N1, portions of the Eastern and Western filaments are also present in this channel, with the Western filament stronger than the Eastern filament.  By comparing the DIRTY map with the DIRTY beam, one can see that S1 is located at the first sidelobe of the Western filament (other sidelobes also are visible in the DIRTY map).  To test whether S1 is an artifact related to this sidelobe, we also show in Figure~\ref{artifact} DIRTY maps with the Western filament subtracted.  As can be seen, S1 becomes much weaker, and is now no stronger than the remaining sidelobes (from the Eastern filament) or noise peaks in the DIRTY map.  This behavior is repeated in all channels where S1 is detectable, resulting in a spatial-kinematic structure for S1 that closely resembles the Western filament (see Fig.~2 of Paper~I).  This exercise clearly demonstrates that S1 is just a sidelobe related to the Western filament.  A similar exercise reveals that N1 is a sidelobe of the Inner and Eastern filaments combined.   Furthermore, as can be seen in Figure~\ref{3pointing}, both N1 and S1 lie outside the main body of molecular gas mapped with the IRAM~30-m telescope; all the other confirmed genuine filaments coincide with ridges in the main body of molecular gas.

\subsubsection{Position-Velocity Diagrams}\label{position-velocity diagrams}
Figure~\ref{pv_diagram} shows position-velocity (PV-) diagrams of all the filaments along radial directions.  As can be seen, the improved continuum subtraction applied to the data in Paper~I combined with the mosaic data has not resulted in any qualitative changes to the spatial-kinematic structures of any of the filaments, except for the detection of the new feature W2.  The PV-diagram along a position angle of $90\degr$ (top right panel) that cuts through the both the inner extent of the Western filament and W2 shows that W2 is not just spatially but also kinematically distinct from the Western filament.

We had hoped that the resulting improvement in sensitivity (permitting also a slight improvement in angular resolution with uniform weighting) over that attained in Paper~I would reveal velocity patterns in the shorter filaments E1, E2, and W1.  Sadly, no such patterns are discernible.  Both here and in the IRAM PdBI map of \citet{sal08a} at a higher sensitivity, E2 can be seen to comprise two peaks that, especially in the latter map, appear be discontinuous in velocity.  The stronger peak may show a velocity gradient, which if real increases in blueshifted velocities at decreasing radii just like the Eastern and Western filaments.

\subsubsection{Spectra and Molecular Gas Masses}\label{spectra and molecular gas masses}
Figure~\ref{spectra} shows spectra of the individual filaments at a velocity resolution of $20 {\rm \ km \ s^{-1}}$.  In Table~\ref{gas_mass}, we list the integrated flux density of each filament as derived from these spectra.  Note that the integrated flux density for the Western filament contains a small contribution from W2.  The quoted uncertainty in the flux density of the Inner filament includes the uncertainty in determining the continuum subtraction level (a minor component of the error budget).  None of the quoted uncertainties in flux densities, however, include any systematic uncertainty in absolute flux calibration that may be as large as $\sim$20$\%$.  We converted the integrated CO(2--1) flux density of each filament to its corresponding mass in molecular hydrogen (H$_2$) gas using the same assumptions as in Paper~I, thus facilitating a direct comparison with the single-dish map of \citet{sal06} (as discussed later in \S\ref{unrecovered emission}).  Because of a trivial mistake in converting from flux density to brightness temperatures, all the values of molecular hydrogen gas masses quoted in Paper~I are overestimated by a factor of ${1/\rm ln~2}=1.44$ (but those numbers also need to be revised because of the imperfect continuum subtraction).  In Table~\ref{gas_mass}, we list the revised values for the molecular hydrogen gas masses of each filament.  The conversion factor used for relating CO(2--1) flux densities to molecular hydrogen gas masses is that believed to be appropriate for Galactic molecular clouds \citep{sol87}; this conversion factor is estimated to be $\sim$5 times too high for the centrally concentrated molecular gas (disks) in ultraluminous infrared galaxies \citep{dow98}.  The appropriate conversion factor for the molecular gas in Per~A is not known.

The filaments have increasingly larger masses with decreasing angular distance from the center along a direction that passes through the eastern filaments (Eastern filament and E1), as well as through the Inner Filament and the western filaments (Western filament and W1).  
The mass of the Inner filament alone is comparable with that of the Eastern and Western filaments combined, and with the mass of molecular gas in normal spiral galaxies.  These three filaments comprise $\sim$$84\%$ of the total mass of all the detected filaments.  Altogether the filaments have a total mass of $(43.0\pm1.1)\times10^{8} {\rm \ M_{\sun}}$, about one-half that measured by single-dish telescopes over the same region mapped in our mosaic observation of $\sim$$1\times10^{10} {\rm \ M_{\sun}}$ \citep{sal06}.  In $\S\ref{unrecovered emission}$, we examine how much of the CO(2--1) emission detected with single-dish telescopes towards the filaments observed has been recovered in our interferometric observations.

\section{DISCUSSION}
\subsection{Radially-Infalling Gas from a X-ray Cooling Flow}\label{Infalling Gas}
In Paper~I, we found that the three spatially-resolved (Inner, Western, and Eastern) filaments detected are radially aligned.  We showed that the velocity pattern of the Western and Eastern filaments can be simply explained as free-fall in the gravitational potential of Perseus~A.  The improved PV-diagrams for the latter two filaments presented in this paper do not change this central conclusion.  

In this paper, we find that E1, E2, and W1 like the three abovementioned filaments also comprise radially-aligned filaments.  Since our observations were made, \citet{sal08a} also have reported that E1 and E2 comprise radially-aligned filaments.  Thus, in all, six distinct cool molecular filaments, three lying east of center and three lying west of center, have now been identified. 
Their radial morphologies suggest that there is no significant transverse velocity component tracing orbital motion, arguing against the idea that some of the cool molecular gas detected may have been captured from a gas-rich galaxy.  The lack of any transverse spatial component provides no support for (but does not necessarily negate) the idea that the cool molecular gas may have been dragged outwards by buoyant X-ray bubbles, as has been proposed to explain transverse H$\alpha$ filaments seen in the outer regions of the nebula \citep{fab03}.

\subsection{Relationship with Optical Emission-Line Nebula}\label{Relationship with Optical Nebula}
\subsubsection{Morphology and Kinematics}
As mentioned in Paper~I, the observed CO(2--1) filaments (apart from N1 and S1, which as shown in $\S\ref{artifacts}$ turn out to be artifacts) are spatially coincident with bright H$\alpha$ features.  In Figure~\ref{co-haf}, we plot contours of the integrated CO(2--1) intensity on a color image of the H$\alpha$$+$N[II] emission from \citet{con01}.  We have adjusted the contrast for the inner$+$western and eastern regions separately to better see the spatial relationship between the molecular and ionized gas in these regions.  The Inner filament is embedded in bright H$\alpha$ emission that surrounds the nucleus.  In both the eastern and western regions, every molecular filament, as well as the new feature W2, is associated with locally bright H$\alpha$ features.  On the other hand, not all equally bright if not brighter H$\alpha$ features, such as those labeled A--D in the middle panel and HII~region in the bottom panel of Figure~\ref{co-haf}, have CO(2--1) counterparts.  As discussed in Paper~I and also \citet{sal06}, all the cool molecular filaments detected have velocities comparable if not identical with their H$\alpha$ counterparts (where measured).

\subsubsection{Luminosity}\label{luminosity}
Both \citet{edg01} and \citet{sal03} found a linear correlation (albeit with considerable scatter) between the globally integrated CO and H$\alpha$ luminosities for the central cD galaxies in putative X-ray cooling-flow clusters.  \citet{sal06} find that this relationship extends to individual regions mapped in Per~A at an angular resolution of $\sim$12\arcsec\ with the IRAM-30~m.  In Figure~\ref{mass-haf-corr}, we plot the integrated CO(2--1) luminosities of the individual filaments in Per~A against their corresponding H$\alpha$ luminosities measured in an area encompassing the lowest CO(2--1) contour of a given filament as seen in Figure~\ref{co-haf}.  The H$\alpha$ emission associated with the Inner filament contains a substantial contribution from the AGN, and so the value plotted constitutes an extreme upper limit on the emission actually associated with this filament (as indicated by the arrow).  The diagonal lines drawn in Figure~\ref{mass-haf-corr} correspond to different ratios between the CO(2--1) and H$\alpha$ luminosities.  As can be seen, there appears to be an approximately linear relationship between the CO(2--1) and H$\alpha$ luminosities: a least-squares fit (omitting the Inner filament) gives a ratio between the CO(2--1) and H$\alpha$ luminosity of $\sim$0.0007 (indicated by the solid line in Fig.~\ref{mass-haf-corr}) with a correlation coefficient of 0.93.  Thus, the CO(2--1) line has a luminosity of about $0.07\%$ that of the H$\alpha$+N[II] line.  \citet{joh07} also have found a linear correlation between the line luminosities of warm (300-400~K) molecular hydrogen and H$\alpha$+N[II], with the rotational line of H$_2$~0--0S(1) having a luminosity of about $3\%$ that of the H$\alpha$+N[II] line.

The existence of a correlation between the CO(2--1) and H$\alpha$ luminosities of the filaments may suggest that the excitation of their cool molecular gas and ionized gas is causally related.  Alternatively, this correlation may support the picture in which the optical emission-line nebula comprises the ionized skins of individual cool giant molecular hydrogen clouds \citep[e.g.,][]{hec89}: more giant molecular clouds result in stronger CO emission, and a larger surface area exposed to ionization radiation and hence stronger H$\alpha$ emission.

On the other hand, not all H$\alpha$ features that are as bright if not brighter than those associated with the CO(2--1) filaments have detectable CO(2--1).  The most obvious example is in the eastern region, where the brightest H$\alpha$ feature visible in Figure~\ref{co-haf} (bottom panel) does not have a CO(2--1) counterpart.  This H$\alpha$ feature has the spectrum of a HII region and also a very blue continuum characteristic of a young and massive star cluster \citep{shi90}.  By contrast, other parts of the optical emission-line nebula so far studied spectroscopically --- including those along the Inner, Western, and Eastern Filaments, as well as E1 and E2 --- have been found to have both low ionization and excitation \citep{sab00}.  In the inner$+$western region of Figure~\ref{co-haf} (middle panel), examples of comparably bright H$\alpha$ features with no associated CO(2--1) emission are labeled A--D.  These particular H$\alpha$ features have not, to the best of our knowledge, been studied spectroscopically in the optical.  We have computed 3$\sigma$ upper limits for the molecular hydrogen gas masses in all these regions assuming an overall linewidth of $60 {\rm \ km \ s^{-1}}$, as found for the shorter filaments E1, E2, and W1.  For comparison, the HII region found by \citet{shi90} has an upper limit for its linewidth of $\sim$$20 {\rm \ km \ s^{-1}}$ at FWHM \citep{sfb90}.   As shown in Figure~\ref{mass-haf-corr}, the features A--D have upper limits in their luminosity ratio of CO(2--1) to H$\alpha$ that is up to a factor of $\sim$5 lower than that derived above for the CO(2--1) filaments.  The HII region has a corresponding upper limit in this ratio that, based on the more sensitive measurements of \citet{sal08a}, is more than an order of magnitude lower.  Optical spectroscopy of the features A--D is needed to determine whether they also are giant HII regions, or whether there is a relatively large scatter in the CO(2--1) to H$\alpha$ luminosities even between different features in the low ionization and excitation optical emission-line nebula.

\subsection{Gravitational Stability of Cool Molecular Filaments}
Although there is abundant evidence for recent star formation in Per~A \citep[][and references therein]{con01}, there is no evidence for star formation in the molecular filaments that we detect.  As mentioned in $\S\ref{Relationship with Optical Nebula}$, optical spectroscopy of their corresponding H$\alpha$ filaments do not show HII-region-like spectra, but instead gas with low ionization and excitation.  \citet{ler08} have shown that nearby normal spiral galaxies exhibit a nearly constant star formation efficiency (star formation rate per unit of molecular gas) of $\sim$$5 \times 10^{-10} {\rm \ yr^{-1}}$.  If the molecular filaments detected in Per~A are forming stars at the same star formation efficiency, their resulting HII regions should have a H$\alpha$ luminosity comparable with that observed for the H$\alpha$ filaments cospatial with these molecular filaments.  Instead, any HII regions in these filaments can contribute at most a small fraction of the observed H$\alpha$ emission \citep[e.g., see Fig.~3 of][]{sab00}, and therefore the star formation efficiency in these molecular filament must be lower than that found in nearby spiral galaxies.

We show below that, in the absence of any external forces or internal means of support beyond thermal pressure, all the molecular filaments detected are gravitationally bound and should collapse on a timescale at most comparable if not much shorter than the inferred dynamical ages of the Eastern and Western filaments.  We then explore what may be preventing these filaments from collapsing to form stars (with an efficiency comparable with that seen in normal spiral galaxies).

\subsubsection{Binding Mass}

For simplicity, we first model each filament as a string of uniform spherical clouds.  The Jeans mass --- the minimum mass required for a uniform spherical cloud to be gravitationally bound if supported only by thermal pressure --- can be expressed as $M_J \simeq 2.4 \times 10^3 \ T({\rm K}) \ r_c({\rm kpc}) {\rm \ M_{\odot}}$ (assuming that all the mass is in the form of molecular hydrogen gas), where $T$ is the gas temperature and $r_c$ the radius of the cloud.  Studies of the global CO(1--0), CO(2--1), and CO(3--2) line ratios in the main body of molecular gas \citep{bri98}, as well as CO(1--0) and CO(2--1) line ratios measured at an angular resolution of $\sim$12\arcsec\ with the IRAM~30-m in the main body of molecular gas as well as towards the outer H$\alpha$ filaments \citep[]{sal08b}, suggest that the CO(2--1) emission is optically thick.  Modeling of the global CO line ratios suggest multi-temperature components at $\sim$10~K and $\sim$170~K \citep{bri98}.  For comparison, the peak brightness temperature measured for any of the filaments is only $\sim$0.6~K, suggesting that either the filaments have much narrower widths than observed or comprise complexes of molecular clouds with an overall small surface filling factor.

Let us start by considering a spherical cloud with $r_c \approx 0.5 {\rm \ kpc}$, which is the upper limit placed on the half-widths of the filaments.  Each of the observed filaments could then comprise (at least geometrically) up to about eight such clouds.  If at the excitation temperature of the CO(2--1) transition of $T \approx 15$~K, then to be gravitationally bound they are required to have masses of at least $M_J \approx 10^4 {\rm \ M_\sun}$; if at $T \approx 170 {\rm \ K}$, then $M_J \approx 10^5 {\rm \ M_\sun}$.  Such Jean masses are many orders of magnitude smaller than the masses of the individual filaments, implying that any such clouds would certainly be gravitationally bound. 

If instead the filaments comprise complexes of (uniform spherical) molecular clouds with sizes comparable to the range seen for Giant Molecular Clouds (GMCs) in our Galaxy of $r_c\approx 5$--50~pc, $M_J \approx 10^2$--$10^3 {\rm \ M_\sun}$ if at $T \approx 15$~K or $M_J \approx 10^3$--$10^4 {\rm \ M_\sun}$ if at $T \approx 170 {\rm \ K}$ (compared with observed masses for Galactic GMCs of about $10^4$--$10^7 {\rm \ M_\sun}$).  Consider first those clouds with $T \approx 15$~K: to give a measured brightness temperature of $\sim$0.3~K as is typically observed for the filaments, such clouds are required to have surface filling factors of $\sim$$2\%$.  If they do not overlap along the line of sight, then for the longest filament such clouds would number about 10 (for $r_c\approx 50$~pc) to $10^3$ (for $r_c\approx 5$~pc); if in the extreme case where the volume filling factor equals the surface filling factor, then the corresponding number of such clouds is about $10^2$ (for $r_c\approx 50$~pc) to $10^5$ (for $r_c\approx 5$~pc).  For comparison, the masses of the longest filaments are 6--7 orders of magnitude larger than the individual Jeans masses of such clouds, implying once again that (the vast majority of) such clouds would most likely be gravitationally bound.  A similar computation for clouds at $T \approx 170$~K leads to the same conclusion.

\subsubsection{Timescale for Gravitational Collapse}\label{timescale gravitational collapse}
The characteristic free-fall time, $\tau_{ff}$, of a uniform spherical gas cloud (in the absence of rotation and any other means of support beyond thermal pressure such as turbulence or magnetic fields) depends only on its mass density, $\tau_{ff} \simeq (2 {\rm \ Myr}) \ ({10^3 {\rm \ cm^{-3}}/n})^{1/2}$, where $n$ is the molecular hydrogen gas number density.  The upper limits placed on the volume of the filaments, together with the unknown filling factor of the molecular gas in these filaments, do not translate into meaningful upper limits for the gas density.  \citet{bri98} find from modeling the global CO(1--0), CO(2--1), and CO(3--2) line ratios in Perseus~A a molecular hydrogen gas density $n \approx 10^3 {\rm \ cm^{-3}}$; i.e., close to the critical density of molecular hydrogen gas for collisional excitation of the CO(2--1) transition.  Such densities are typical for the bulk of GMCs in our Galaxy.  If the filaments, modeled as cylinders, are uniformly filled with gas at such densities, then they would have radial cross-sections of just tens of parsecs or smaller.  The actual gas densities may be higher, implying an upper limit for the free-fall time of $\sim$$10^6 {\rm \ yrs}$.  For comparison, the Eastern and Western filaments have dynamical ages of about 20--30~Myr (with a likely uncertainty of roughly $\pm 2$), about an order of magnitude longer than the free-fall time.

The filaments as modeled above (string or complex of uniform spherical gas clouds) will undergo gravitational collapse at the free-fall time if the sound crossing time is longer.  The sound crossing time can be expressed as $\tau_{s} \simeq (25 {\rm \ Myr}) \ r_c(\rm pc) \ c_s^{-1}(\rm km \ s^{-1})^{-1}$, where $c_s$ is the sound speed ($\sim$$1 {\rm \ km \ s^{-1}}$ at $T \approx 100$~K).  Even for values as small as $r_c \gtrsim 10$~pc, $\tau_{s} \gtrsim 10^7$--$10^8$~yrs, much longer than the free-fall time.  The same is true for GMCs in our Galaxy, which have sound crossing times of $\gtrsim$$10^7$~yrs.  The estimated ages of GMCs in our Galaxy (a highly controversial subject) ranges from $\sim$$10^6$~yr \citep[][and references therein]{har01} to $\sim$$10^7$~yr \citep[][and references therein]{tas04,mou06}.  The former is consistent with gravitational collapse at the free-fall time, whereas the latter requires GMCs to possess other internal means of support beyond thermal pressure.  The Eastern and Western filaments therefore have dynamical ages at if not significantly beyond the inferred maximum ages of GMCs in our Galaxy.

\subsubsection{Tidal Shear}
We first consider whether the tidal shear exerted by Perseus~A on the filaments along their (radially aligned) lengths can prevent them from collapsing (at least in this direction).  We use the same analytical expression for the gravitational potential of Perseus~A as in Paper~I:
\\
\begin{equation}
\phi(r) = - {G M \over {r+a}} \ ,
\end{equation}
\\
where $G$ is the gravitational constant, $M$ the mass of the galaxy, $r$ the radius from center, and $a = 6.8 {\rm \ kpc}$ for Perseus~A (see Eq.~(5) of Paper~I).  For simplicity, we assume that the filaments can be approximated as uniform cylinders with masses $M_{cy}$ and lengths $l_{cy}$.  The (outwards) tidal force experienced by a test particle along the symmetry axis of such a cylinder at its ends (i.e., distance ${1\over2} l_{cy}$ from center) is therefore
\\
\begin{equation}
F_t(r) \approx G M m {l_{cy}  \over (r + a)^3} \ ,
\end{equation}
\\
where $m$ is the mass of the test particle.  The (inward) gravitational force exerted by the filament itself on that test particle is given by
\\
\begin{equation}
F_{cy}(l_{cy}) = 2 {G M_{cy} m \over r_{cy}^2} [ 1 + {r_{cy} \over l_{cy}}  - \sqrt{1 + ({r_{cy} \over l_{cy}})^2} \ ] \ ,
\end{equation}
\\
where $r_{cy}$ is the radius of the cylinder.  For thin cylinders where $r_{cy} \ll l_{cy}$,
\\
\begin{equation}
F_{cy}(l_{cy}) \approx 2 {G M_{cy} m \over r_{cy} \ l_{cy}} \ .
\end{equation}
The ratio of the outward tidal force exerted by Perseus~A to the inward gravitational force exerted by the cylinder at its ends is therefore:
\\
\begin{equation}
{F_t(r) \over F_{cy}(l_{cy})} \approx {1 \over 2} {M \over M_{cy}} {r_{cy} \ l_{cy}^2 \over (r + a)^3} \ \ \ , \ {\rm for} \ r_{cy} \ll l_{cy} \ .
\end{equation}

Let us consider the Eastern filament, which has a mass $M_{cy} \approx 5 \times 10^8 {\rm \ M_\sun}$ and, assuming an inclination of $\sim$$47\degr$ (Paper~I), a deprojected length $l_{cy} \approx 3.3$~kpc located at a deprojected radius $r \approx 9.4 {\rm \ kpc}$.  With $M \approx 3.4 \times 10^{11} {\rm \ M_\sun}$ for Perseus~A (see Paper~I), and defining $r_{cy} = k \ l_{cy}$ where $k \ll 1$, the ratio $F_t(r) / F_c(l_c) \approx 2.7 \ k$.   We do not resolve the Eastern filament in width (i.e., $r_{cy} \lesssim 0.5$~kpc), implying that $k \lesssim 0.15$ and so $F_t(r) / F_{cy}(l_{cy}) \lesssim 0.4$.  Thus, the outward tidal force is at best comparable with the inward gravitational force of the filament itself, and much less if the filament is much thinner than the upper limit imposed by our observation.

The Western filament has a mass $M_{cy} \approx 1 \times 10^9 {\rm \ M_\sun}$ and, assuming an inclination of $\sim$$40\degr$, a deprojected length $l_{cy} \approx 3.6$~kpc located at a deprojected radius $r \approx 5.5 {\rm \ kpc}$.  Once again we do not resolve this filament in width, implying that $k \lesssim 0.14$ and so $F_t(r) / F_{cy}(l_{cy}) \lesssim 0.6$.  Thus, just like the Eastern filament, the outward tidal force is at best comparable with the inward gravitational force of the filament itself.

Assuming an inclination of $\sim$$45\degr$, we find $F_t(r) / F_{cy}(l_{cy}) \lesssim 0.1$ for E1, $F_t(r) / F_{cy}(l_{cy}) \lesssim 0.2$ for E2, and $F_t(r) / F_{cy}(l_{cy}) \lesssim 0.6$ for W1.  For these shorter filaments located at relatively large distances from center, the outward tidal force is weaker than if not negligible compared with the inward gravitational force of the filaments themselves.

The Inner filament exhibits more complex kinematics than the Eastern and Western filaments, and as explained in Paper~I may comprise two separate filaments.  Assuming that it is a single filament oriented at an inclination of $\sim$$45\degr$, its center would then be located at $r \approx 1.5 {\rm \ kpc}$.  With a mass $M_{cy} \approx 2 \times 10^9 {\rm \ M_\sun}$ and deprojected length $l_{cy} \approx 3.7$~kpc, $k \lesssim 0.14$ and hence $F_t(r) / F_{cy}(l_{cy}) \lesssim 1.1$.  Thus, even for the Inner filament where the tidal shear is strongest, the outward tidal force is at best comparable with the inward gravitational force of the filament itself.

Tidal shear becomes negligible if each of the filaments comprise a string or complex of GMCs, modeled here as uniform spherical clouds with densities of at least $n \approx 10^3 {\rm \ cm^{-3}}$.  Consider a single such cloud located at the midpoint of the filament (so that gravitational forces from clouds on opposing sides of the filament cancel) with mass $M_c$ and radius $r_c$.  The gravitational force on a test particle with mass $m$ at a given radius $b$ in the cloud is given by:
\\
\begin{equation}
F_c(b) = {G M_c m b \over r_c^3}  \ ,
\end{equation}
\\
The ratio of the outward tidal force exerted by Perseus~A (replacing $l_{cy}$ with $2b$ in Eq.~(2)) to the inward gravitational force exerted by the sphere is then given by:
\\
\begin{equation}
{F_t(r) \over F_c(b)} \approx {1 \over 2} {M \over M_c} {r_c^3 \over (r + a)^3}  \ .
\end{equation}
\\
The largest and most massive GMCs in our Galaxy have $M_c \approx 10^7 {\rm \ M_\sun}$ and $r_c \approx 50$~pc (and thus $n \approx 10^3 {\rm \ cm^{-3}}$), giving $F_t(r)/F_c(b) \lesssim 10^{-3}$ at the location of the different filaments.  As $n \propto {M_c / r_c^3}$, this result holds for GMCs of any dimensions provided they all have the same densities as considered here.

As shown in Paper~I, the spatial-kinematic distribution of the Eastern and Western filaments can be modeled as free-fall in the gravitational potential of Perseus~A provided that the mass of this galaxy is about a factor of $\sim$2 smaller than that inferred by \citet{smi90}.  Equivalently, the measured (deprojected) velocity gradients of the Eastern and Western filaments are shallower than those predicted for free-fall if Perseus~A has the mass inferred by \citet{smi90}.  We mentioned in Paper~I that our simple model neglected a number of effects, in particular the outward pressure exerted by the X-ray gas on the infalling filaments.  Here we note that, over time, the self-gravity of the filaments also can reduce any intrinsic velocity gradient along their lengths.

\subsubsection{Turbulent Support}\label{Turbulent Support}
If the molecular filaments are supported by turbulence, then the required turbulent velocities can be estimated from the virial theorem expressed as $v \simeq (GM_s/2r_s)^{1/2}$ for a uniform sphere, where $v$ is the average particle velocity.  As an illustration, take the least massive and shortest filament W1, which has a mass of $\sim$$1 \times 10^8 {\rm \ M_\sun}$ enclosed in $r_s \lesssim 0.5$~kpc.  The required velocity of the molecular hydrogen gas particles to support W1 against collapse is $\gtrsim 21 {\rm \ km \ s^{-1}}$ (note that thermal velocities are $\lesssim 1 {\rm \ km \ s^{-1}}$ for temperatures $\lesssim$100~K).  This is about a factor of $\sim$2 smaller than the CO(2--1) linewidth of W1 (Table~\ref{gas_mass}), indicating that this filament may be supported by turbulence.  An upper limit of up to a factor of several higher turbulent velocity is required to support the other more massive filaments against collapse, which in all cases is smaller than their observed linewidths.  Observations at higher angular resolutions are required to measure the intrinsic linewidths of the filaments (i.e., omitting systematic motions) or their constituent GMCs so as to determine their true turbulent velocities.  At the present time, we cannot rule out the possibility that the filaments are supported by turbulence, generated perhaps in the process of condensing from the X-ray cooling flow.

Turbulence normally cascades to ever smaller spatial scales, and dissipates on a timescale shorter than the crossing time at the maximum turbulent velocity \citep[e.g.,][]{sto98}.  For the derived parameters of W1, this turbulent decay timescale is at most $\sim$$10^7$~yrs.  The other filaments also have similar turbulent crossing times, comparable with if not shorter than the dynamical ages of the Eastern and Western filaments.  Thus, it may be that a source is required to at least partially regenerate turbulence over the inferred lifetimes of the Eastern and Western filaments.
GMCs in our Galaxy also have supersonic linewidths, typically several $\rm km \ s^{-1}$; how such turbulence is maintained is not understood, even though these GMCs have lifetimes that are at most comparable with the dynamical ages of the Eastern and Western filaments.

\subsubsection{Magnetic Support}
If the molecular filaments are supported by magnetic fields, we can make the following simplifying assumptions to roughly estimate the required field strengths.  In a uniform magnetic field, the filaments (or GMCs) will collapse along the field lines but be (temporarily) prevented from collapsing perpendicular to the field lines (by collisions with ions) to form thin sheets.  The magnetic flux, $\Phi$, required to support a uniform sheet with a given mass, $M_{sh}$, against its own gravity is given by $\Phi / M_{sh} = 2 \pi \sqrt{G}$ \citep{nak78}.  For example, if the Western filament comprises a sheet with a deprojected diameter of 3.6~kpc and a mass $M_{sh} \approx 1 \times 10^9 {\rm \ M_{\sun}}$, it is required to have a magnetic field strength of at least $\sim$$30 {\rm \ \mu G}$ to prevent its (immediate) collapse.  The other filaments require minimum magnetic field strengths in the range $\sim$10--$60 {\rm \ \mu G}$.  This is towards the low end of the range measured for the magnetic field strengths in Galactic GMCs \citep{cru99}.  \citet{fab08} suggest that the thread-like H$\alpha$ filaments are supported by magnetic fields with strengths of a few tens of $\mu$G, comparable with the values estimated here to support the molecular filaments against collapse (in the absence of other means of nonthermal support such as rotation and turbulence).

Ambipolar diffusion \citep{mes56} will eventually cause the filaments (or their constituent complexes of GMCs) to collapse across the magnetic field lines, as is thought to be the case for GMCs in our Galaxy.  The ambipolar diffusion timescale is given by $\tau_{AD} \simeq (\tau_{ff})^2/\tau_{ni}$ \citep{mou79,mou82}, where $\tau_{ni}$ is the neutral-ion collision time, in the situation where $M_{sh}/\Phi \ll 1$ \citep{cio01}.  For GMCs in our Galaxy, $\tau_{AD} \approx 1$--10~Myr depending on $M_{sh}/\Phi$ in specific regions of the cloud.  For the Eastern and Western filaments to resist ambipolar diffusion, they also are required to have neutral-ion (and/or neutral grain) collision times comparable with if not significantly longer than that inferred for Galactic GMCs.

\subsubsection{Star Formation}
Although there is no evidence for star formation in the molecular filaments that we detect, there is evidence for recent and widespread star formation in Per~A \citep[][and references therein]{con01}, including the HII region mentioned in $\S\ref{Relationship with Optical Nebula}$ discovered by \citet{shi90}.  
Assuming that any associated CO(2--1) emission has an overall linewidth of $\sim$$20 {\rm \ km \ s^{-1}}$ (see $\S\ref{luminosity}$), we place a 3$\sigma$ upper limit of $3 \times 10^7 {\rm \ M_\sun}$ on its associated mass in molecular hydrogen gas (a smaller linewidth would result in an even smaller upper limit).  An even more stringent upper limit (assuming the same line properties) of $4 \times 10^6 {\rm \ M_\sun}$ can be placed on any associated molecular hydrogen gas from the more sensitive observation of \citet{sal08a} with the IRAM~PdBI.  For comparison, \citet{shi90} infer a mass for the stellar cluster of $\sim$$5 \times 10^6 {\rm \ M_\sun}$ assuming a Salpeter IMF.  If $\sim$$1\%$ of the pre-existing molecular hydrogen gas has been converted to stars in this cluster as is typically inferred for GMCs in our Galaxy, one would have expected to find $\sim$$5 \times 10^8 {\rm \ M_\sun}$ of molecular hydrogen gas associated with this cluster.  Unless this gas has been largely dissociated or widely dispersed, or the assumption of a Salpeter IMF invalid (i.e., top-heavy IMF), the star-formation efficiency associated with this cluster must have been extraordinarily high; i.e., nearly all the molecular hydrogen gas has been converted into stars.

\subsection{Unrecovered Single-Dish CO(2--1) Emission}\label{unrecovered emission}
As mentioned in $\S\ref{spectra and molecular gas masses}$, we recovered only about one-half of the CO(2--1) emission measured by single-dish telescopes for the main body of the molecular gas in Per~A.  In this region, about half of the CO(2--1) emission must therefore be quite smoothly distributed on angular scales larger than $\sim$5\arcsec--10\arcsec, and hence resolved out in our interferometric observations, when observed at velocity resolutions of $\sim$$20 {\rm \ km \ s^{-1}}$.

\citet{sal06} list in their Table~3 the integrated CO(2--1) intensities at locations in their single-dish IRAM~30-m map where the S/N ratio $\geq 4$.  We have extracted from this table those locations that best overlap with the CO(2--1) filaments detected in our observations.  In Figure~\ref{salome} (left panel), we plot circles showing the FWHM of the IRAM~30-m primary beam centered at these locations on our integrated CO(2--1) intensity map of Figure~\ref{mom-all}.  To compare with the IRAM~30-m measurements, we first convolved our channel maps to the same angular resolution as the IRAM~30-map (i.e., with a Gaussian beam of size 12\arcsec\ at FWHM), and then computed the integrated flux densities at the  same locations over the velocity range $-200 {\rm \ km \ s^{-1}}$ to $+200 {\rm \ km \ s^{-1}}$.  The integrated flux densities measured in the IRAM~30-map versus those measured in our SMA maps are plotted in the graph shown also in Figure~\ref{salome} (right panel).

As can be seen, there is an overall correspondence between the CO(2--1) emission detected in our interferometric and that detected in the single-dish observations.  We recovered (within measurement uncertainties) all the CO(2-1) emission in the eastern region, comprising the Eastern filament, E1, and E2.  \citet{sal08a} reached the same conclusion for their map of this region using the IRAM~PdBI, even though their observations do not have baselines as short as ours.  We also recovered all the CO(2--1) emission in the inner region.  The emission that we measured in the western region, however, is systematically below that measured in the single-dish observations.  In this region, we recover less than half of the CO(2--1) emission present in the IRAM~30-m map.  To recover the remaining CO(2--1) emission, we plan to use the sub-compact configuration of the SMA to provide shorter baselines than already available in our observations thus far.

\section{SUMMARY AND CONCLUSIONS}
Our (Paper~I) previous interferometric CO(2--1) observation of Per~A spanned its central region of radius $\sim$10~kpc.  At a spatial resolution of $\sim$1~kpc, this observation revealed three radial filaments lying approximately east-west, together with a number of other features that were at best poorly resolved.  In this manuscript, we reported follow-up interferometric CO(2--1) observations that extend our coverage in the east-west direction to span the entire main body of the cool molecular gas, which in single-dish maps can be traced out to $\sim$14~kpc east and west of center.  Our results showed that:

\begin{itemize}

\item[1.]  No new features were detected beyond the region previous mapped, with the most distant features detectable in our maps lying $\sim$10~kpc east (E1) and $\sim$8~kpc west (W1) of center.  The two CO(2--1) features referred to in Paper~I as N1 and S1 lying north and south of the center are sidelobe artifacts.   All the other features are confirmed to be real.  A previously unrecognized feature with distinct kinematics labeled W2 (Fig.~\ref{mom-all}) is detected partially blended with the Western Filament lying south about halfway along the length of this filament.

\item[2.]  With our improved sensitivity and angular resolution from combining the present and previous observations, all the CO(2--1) features (except W2) can now be seen to comprise filaments that (within measurement uncertainties) are radially aligned with respect to the center of Per~A.  Six distinct filaments (Inner Filament, Eastern Filament, E1, E2, Western Filament, W1) can now be identified.  Our sensitivity and/or angular resolution is not sufficient to discern a velocity pattern in the E1, E2, and W1 filaments.  The newly recognized feature W2 is not spatially resolved.

\item[3.]  All the detected filaments exhibit a roughly constant ratio between their CO(2--1) and H$\alpha$ luminosities of $\sim$$10^{-3}$.  On the other hand, not all equally if not more luminous H$\alpha$ features in the area observed --- including the HII region discovered by \citet{shi90} that lies just "downstream" of E2 --- have detectable CO(2--1) emission.  The upper limits on their CO(2--1) to H$\alpha$ luminosities is a factor of $\sim$5 lower that measured for the CO(2--1) filaments.  An even more stringent upper limit for this ratio of an order of magnitude lower than that derived for the CO(2--1) filaments can be placed for the HII region from the more sensitive observation of \citet{sal08a}.

\item[4.]  About one-half of the CO(2--1) emission detected in single-dish observations with the IRAM~30-m telescope \citep{sal06} from the main body of the molecular gas was recovered in our interferometric observations with the SMA.  Towards the inner and eastern regions, all (or the bulk) of the CO(2--1) emission was recovered.  On the other hand, towards the western region, less than half of the CO(2--1) emission was recovered.

\end{itemize}

We interpreted our results in the following manner:

\begin{itemize}

\item[1.]  The radial alignment of E1, E2, and W1 provides further support for our argument that all the cool molecular gas detected is most likely deposited by a X-ray cooling flow.
The lack of any transverse component in the cool molecular filaments argues against (but does not negate) the idea that this gas may have been dragged outwards by buoyant X-ray bubbles \citep[e.g.,][]{rev08,sal08a}, as has been proposed to explain transverse H$\alpha$ filaments seen in the outer regions of the nebula \citep{fab03}.  In addition, there are no detectable X-ray bubbles in the directions of these filaments.

\item[2.]  The existence of a correlation between the CO(2--1) and H$\alpha$ luminosities for all the filaments detected may suggest that the excitation of their cool molecular and ionized gas is causally related.  Alternatively, this dependence may support the picture in which the H$\alpha$ gas comprises the ionized skins of the cool molecular gas filaments \citep[e.g.,][]{hec89}.

\item[3.]  If the molecular filaments are forming stars, then their star formation efficiency must be much lower than the nearly constant star formation efficiency found for molecular clouds in nearby normal spiral galaxies. 

\item[4.]  The lack of cool molecular gas associated with the giant HII region discovered by \citet{shi90} implies that, unless this gas has been largely dissociated or widely dispersed, or the stellar IMF distinctly non-Salpeter (i.e., top-heavy IMF), the star formation in this region must have been extraordinarily efficient (i.e., essentially all the molecular gas must have been converted to stars).  Optical spectroscopy of the features labeled A--D in Figure~\ref{co-haf} is needed to determine whether they also are giant HII regions, or whether there is a relatively large scatter in the CO(2--1) to H$\alpha$ luminosities even between different features in the low ionization and excitation optical emission-line nebula.

\end{itemize}

Finally, we provided supporting arguments demonstrating that:

\begin{itemize}

\item[1.]  The outward tidal force exerted by Perseus~A on the individual filaments is at best comparable to, and for the shorter filaments much weaker than, the inward gravitational force exerted by the filaments themselves.  If comprising a string or complex of GMCs, all the individual GMCs are gravitationally bound.  Such GMCs, if supported only by thermal pressure, should collapse on timescales no longer than $\sim$$10^6$~yrs.  By comparison, the inferred dynamical ages of the Eastern and Western filaments are an order of magnitude longer, $\sim$$10^7$~yrs (Paper~I).

\item[2.]  If the cool molecular filaments are supported by turbulence, non-thermal velocities of a few to several tens of ${\rm km \ s^{-1}}$ are required.  This is smaller than the observed CO(2--1) linewidths of all the filaments at a spatial resolution of $\sim$1~kpc, although observations at higher angular resolutions are required to measure the true intrinsic linewidths of these filaments (or their constituent GMC complexes) and hence their actual turbulent velocities.  Such turbulence should decay on a timescale no longer than $\sim$$10^7$~yrs, comparable with if not shorter than the dynamical ages of the Eastern and Western filaments.  The filaments may therefore be supported by turbulence, generated perhaps in the process of condensing from the X-ray cooling flow, although a source may be required to maintain this turbulence over the inferred lifetimes of the Eastern and Western filaments.

\item[3.]  Magnetic fields with strengths of at least several $\sim$$10~\mu$G are required to support (perpendicular to the field lines) the cool molecular filaments against their own gravity.  This is a comparable with the estimated magnetic field strengths required to support the thread-like H$\alpha$ filaments \citep{fab08}.  To resist ambipolar diffusion for timescales of $\sim$$10^7$~yrs (in the absence of any other means of support beyond thermal pressure), magnetic field strengths of this order and neutral-ion (and/or neutral grain) collision times similar to that in Galactic GMCs are required.

\end{itemize}

\begin{figure}
\begin{center}
\includegraphics[width=160mm]{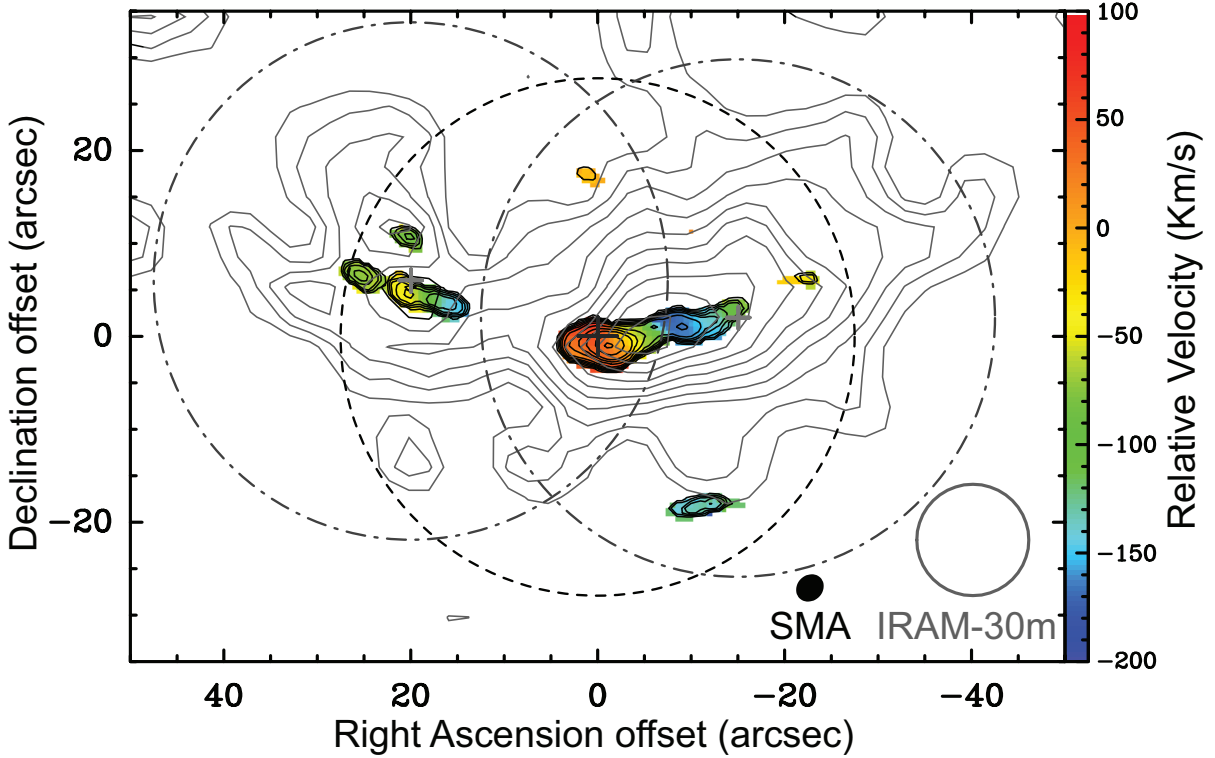}
\vspace{0cm}
\caption{\small Single-dish IRAM-30~m measurements of the integrated CO(2--1) line intensity in light contours \citep[from][]{sal06} overlaid on our previous \citep[from][]{lim08} SMA map of the integrated CO(2--1) line intensity in dark contours and velocity field in color corrected for a small mistake in subtracting the nuclear continuum source (see $\S\ref{observations and data reduction}$ of text).  Dark contour levels are plotted at 3, 6, 9, 12, 15, 20, 30, 40, 50, 70, and $90 \times 0.32{\rm \ Jy \ km \ s^{-1}}$.  The dash circle corresponds to the FWHM of the SMA primary beam centered on the center of Per~A, which is marked by the cross at the origin.  This is the only region covered in our previous observation.  The two dash-dot circles correspond to the FWHM of the SMA primary beams centered approximately on the Eastern and Western filaments as marked by the two crosses to the east and west of center respectively.  Our present observation covers all three regions, thus spanning the main body of cool molecular gas in Per~A.  The angular resolutions of the single-dish and interferometric measurements are shown in the lower right corner.  The larger circle corresponds to the FWHM of the IRAM-30~m primary beam, and the smaller filled ellipse to the FWHM of the SMA synthesized beam.\label{3pointing}}
\end{center}
\end{figure}

\begin{figure}
\begin{center}

\includegraphics[width=75mm]{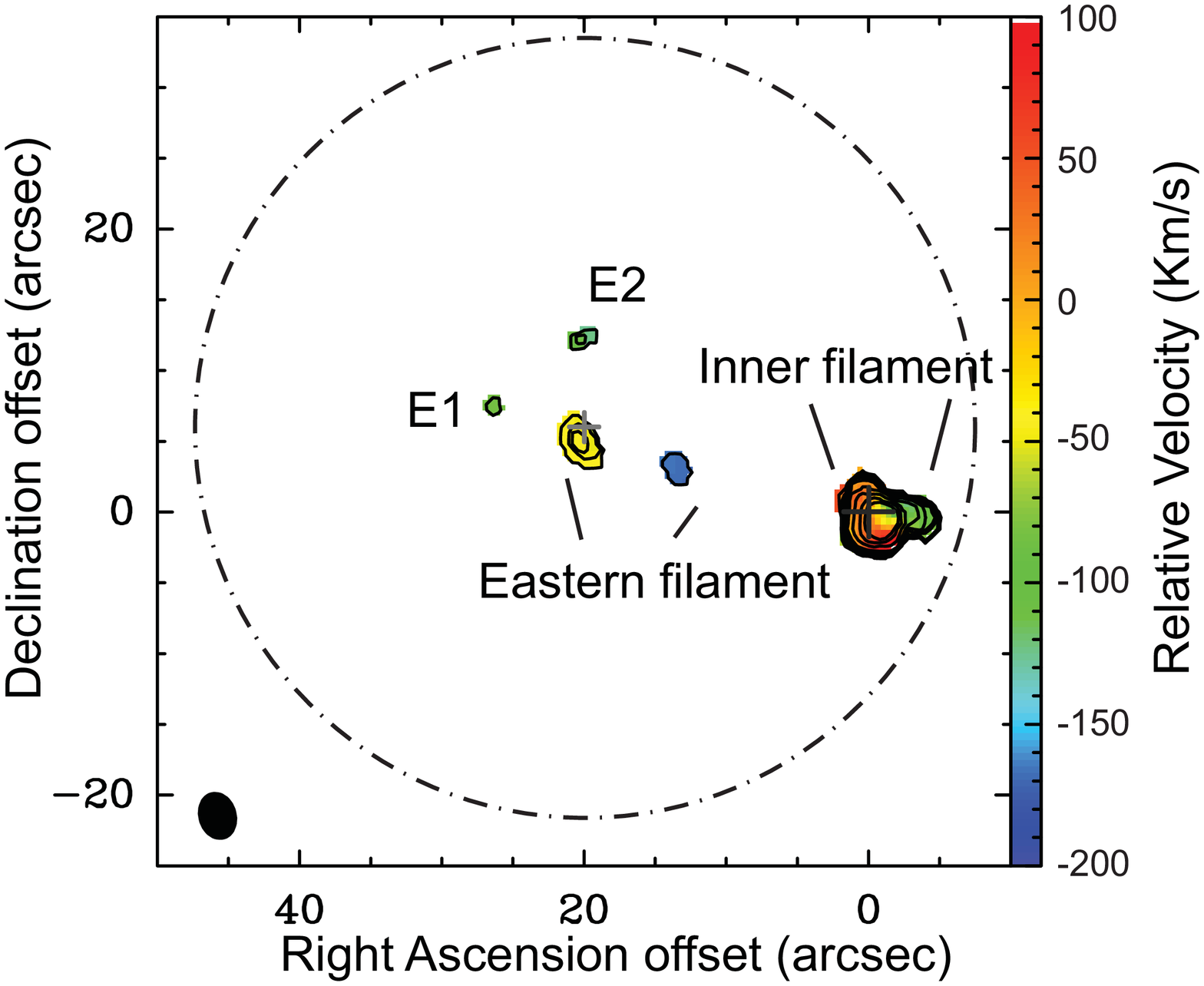} \\
\includegraphics[width=75mm]{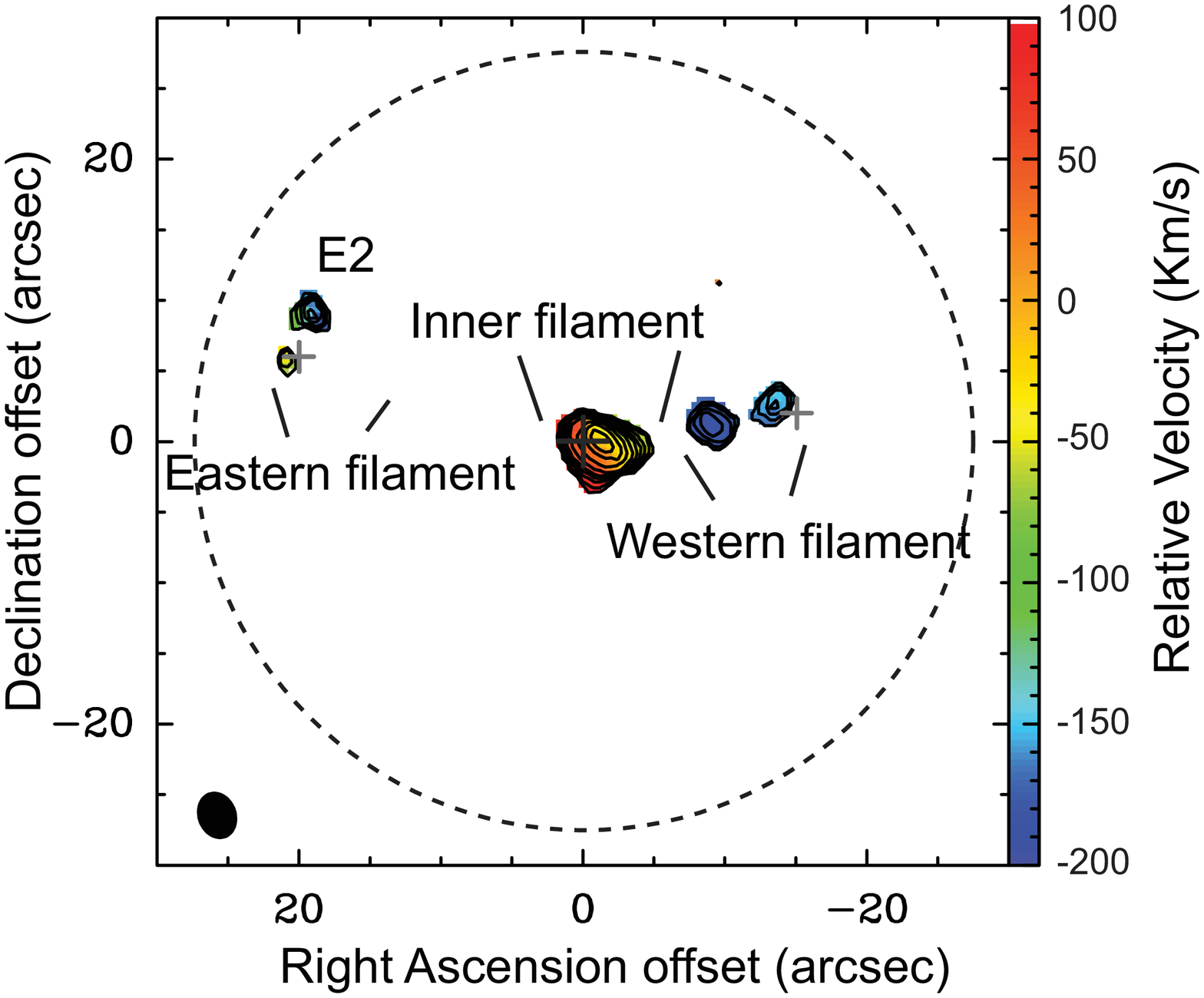} \\
\includegraphics[width=75mm]{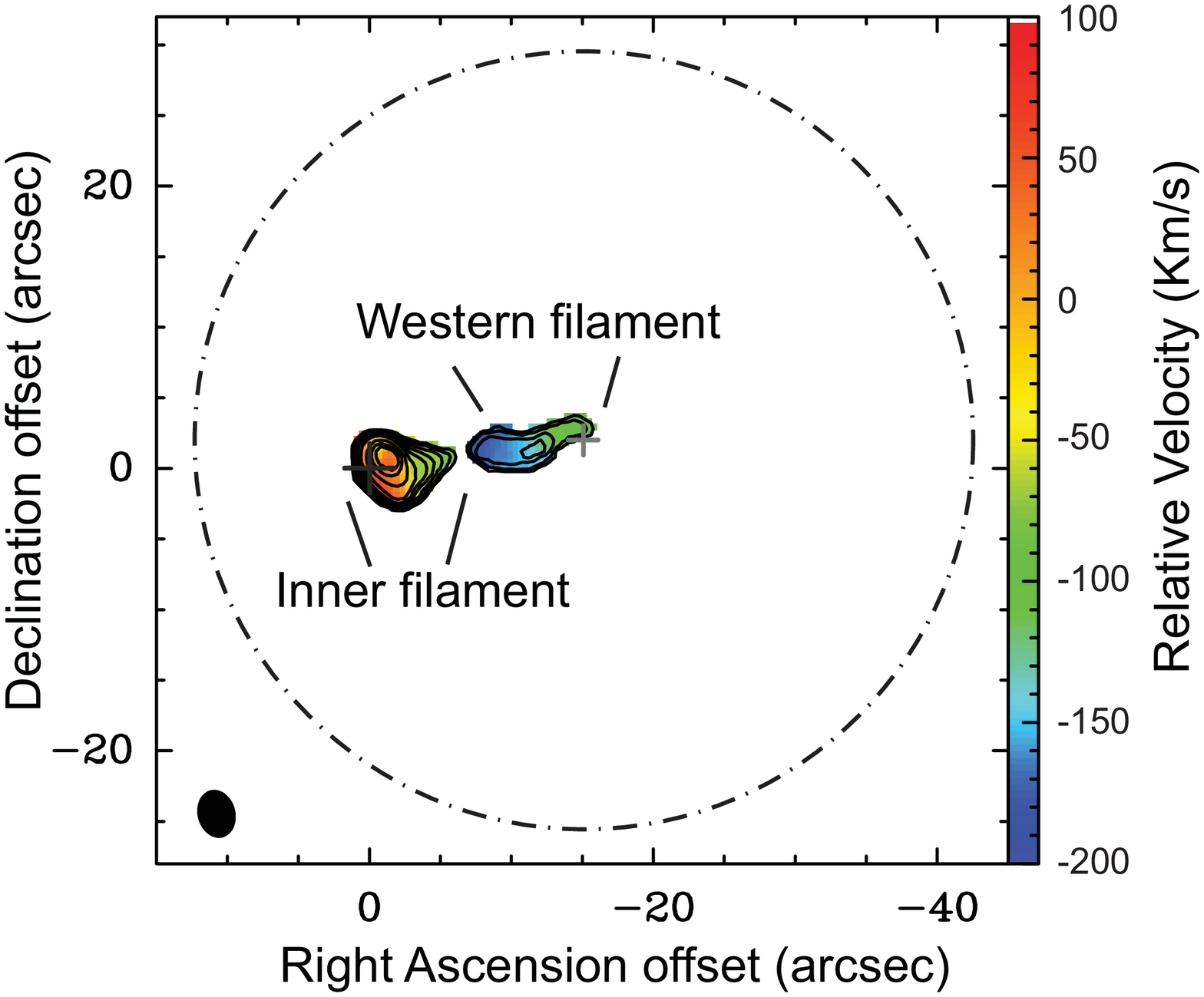}
\vspace{-0.5cm}
{\renewcommand{\baselinestretch}{0.9}
\caption{\small Contours of the integrated CO(2--1) intensity and color maps of the intensity-weighted mean velocity measured with respect to the systemic heliocentric velocity of $5264 \pm 11 {\rm \ km \ s^{-1}}$ \citep{huc99} for the eastern (top panel), central (middle panel), and western pointings (bottom panel).  Contour levels for the central pointing are plotted at 3, 6, 9, 12, 15, 20, 30, 40, and $50 \times 0.6 {\rm \ Jy \ km \ s^{-1}}$, and for the eastern and western pointings at 3, 6, 9, 12, 15, 20, 30, 40, and $50 \times 0.5 {\rm \ Jy \ km \ s^{-1}}$.   The dash and dash-dot circles correspond to the primary beams of the SMA for each pointing.  The synthesized beams for each pointing are plotted at the lower left corner of each panel.  These moment maps were made by first Hanning smoothing their corresponding channel maps by $140 {\rm \ km \ s^{-1}}$, $100 {\rm \ km \ s^{-1}}$, and $60 {\rm \ km \ s^{-1}}$ (at full -width zero intensity) for the eastern, central, and western pointings, and then discarding pixels in the channel maps with intensities less than $1.4\sigma$, $1.7\sigma$, $2\sigma$ for these three pointings respectively. \label{new_mom}}}
\end{center}
\end{figure}

\begin{figure}
\begin{center}
\includegraphics[width=110mm]{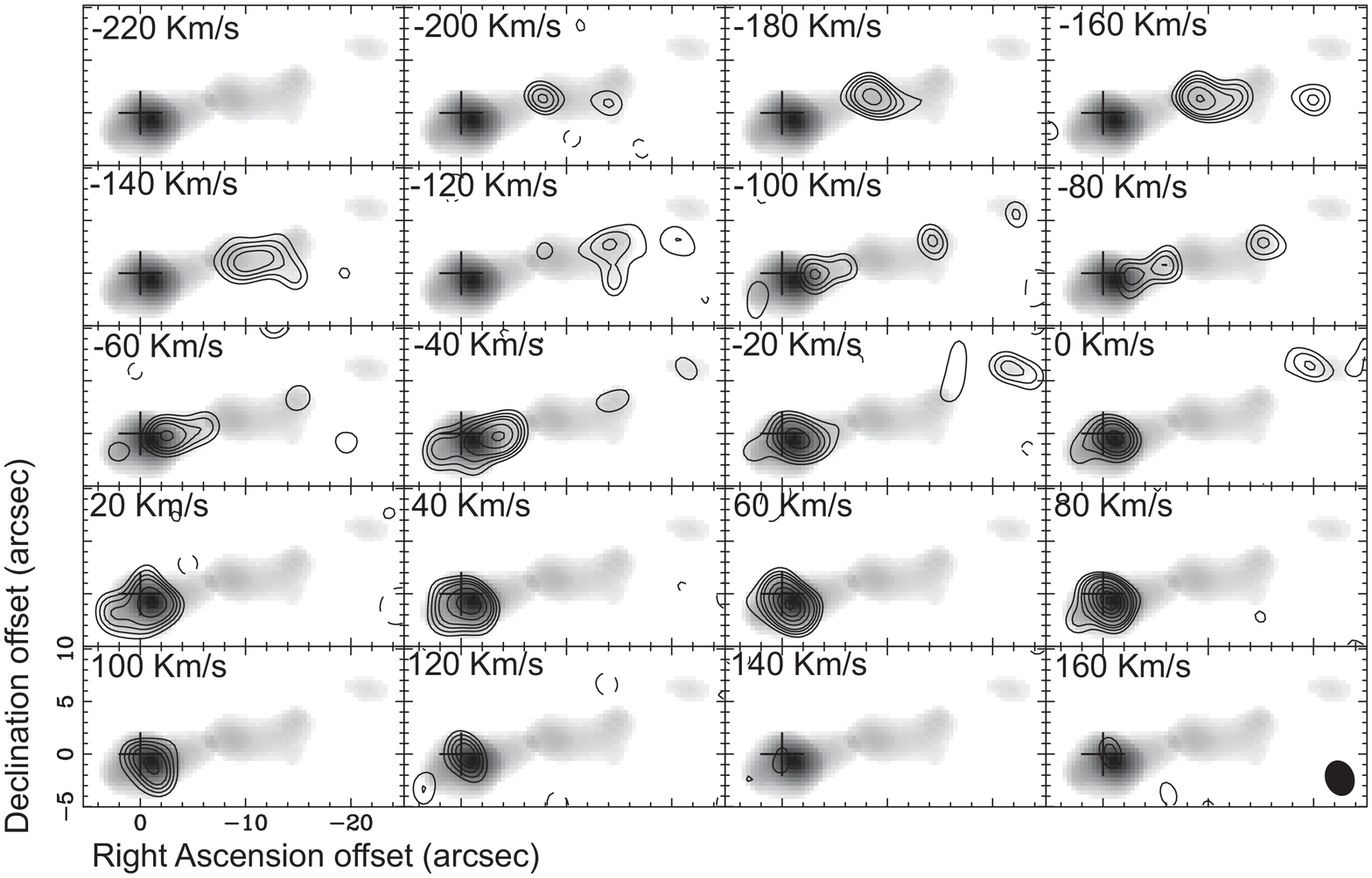} \\
\vspace{0.5cm}
\includegraphics[width=110mm]{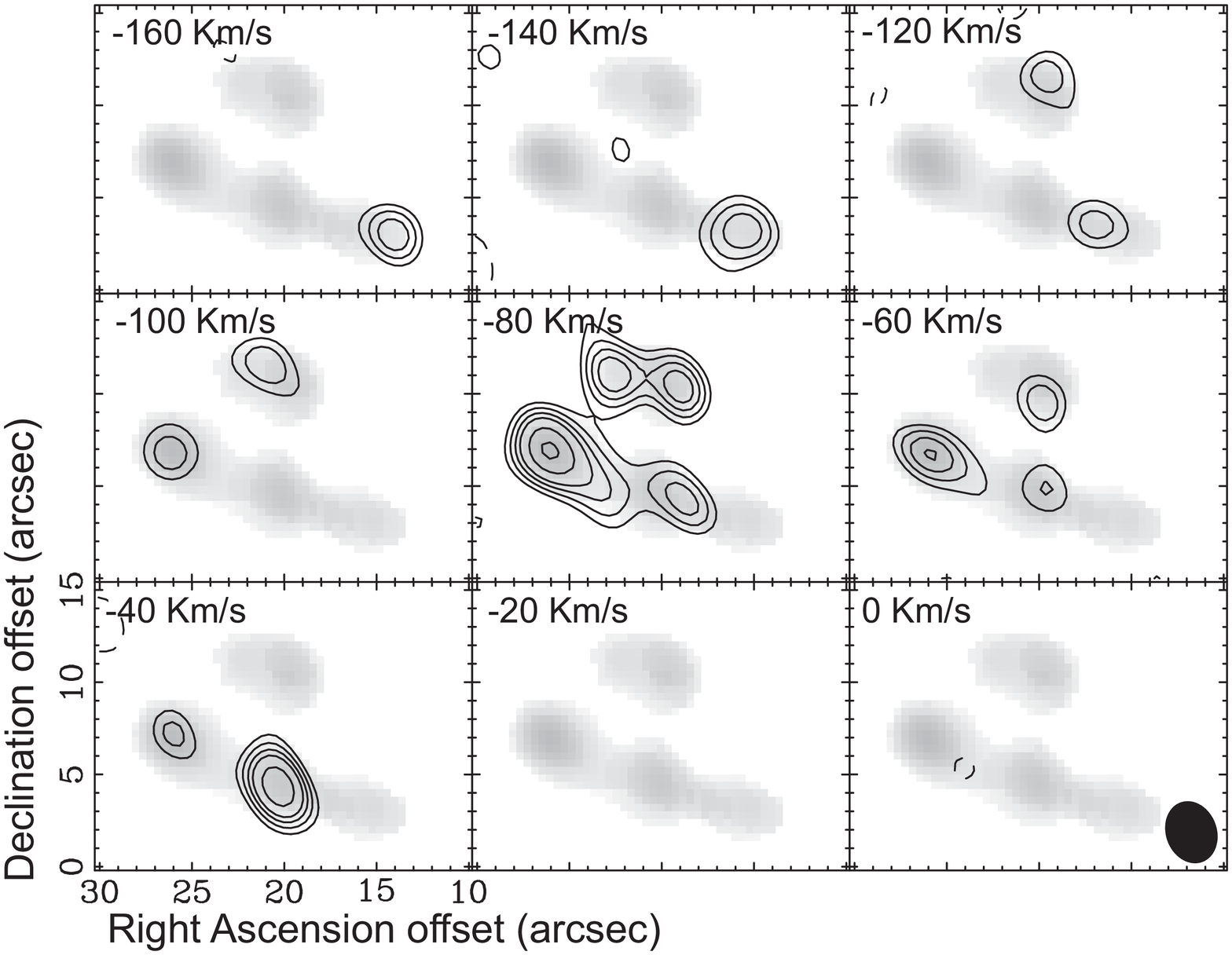} \\
\caption{Channel maps showing contours of CO(2-1) intensity with the velocity indicated in each panel measured with respect to the systemic heliocentric velocity.  The inner together with the western regions (upper panel) and eastern region (lower panel) are shown separately.  Contour levels are plotted at -3 (dashed), 3, 4, 5, 6, 8, 10, 12, 14, and $16 \times 15 {\rm \ mJy \ beam^{-1}}$ for the inner and western regions, and -3 (dashed), 3, 4, 5, 6, 8, and $10 \times 18 {\rm \ mJy \ beam^{-1}}$ for the eastern region.  The gray background corresponds to the integrated CO(2-1) intensity after primary beam correction as shown in Figure~\ref{mom-all}.  The position of the nuclear continuum source in Per~A is indicated by a cross.  The synthesized beam is shown as an ellipse at the lower right corner of the lower right panels for the two regions.  \label{channel-maps}}
\end{center}
\end{figure}

\begin{figure}
\begin{center}
\includegraphics[width=120mm]{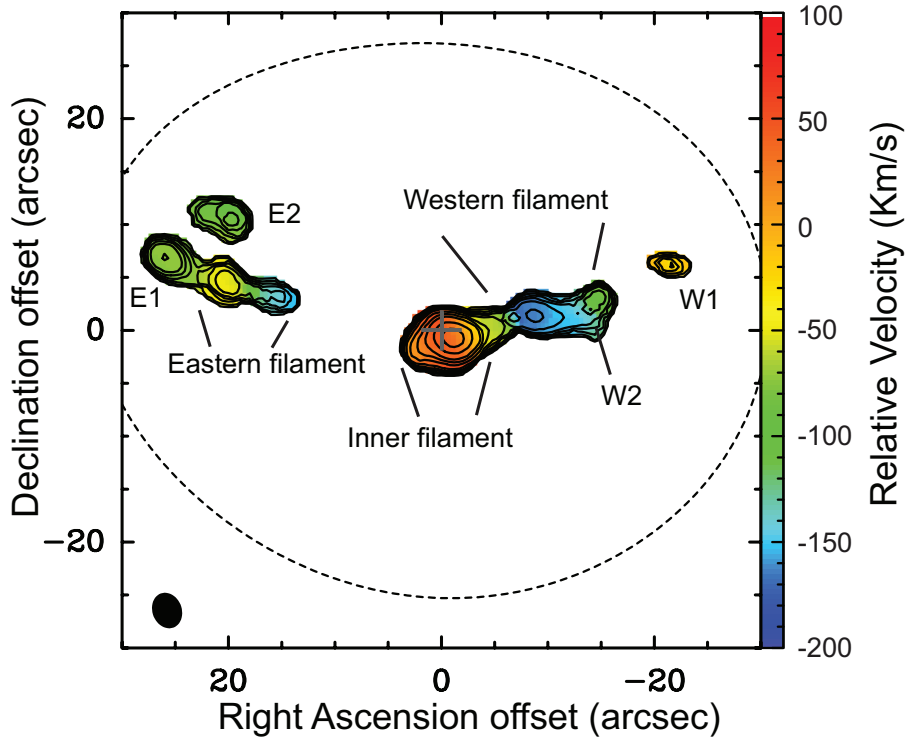}
\caption{Contours of the integrated CO(2--1) intensity and a color map of the intensity-weighted mean velocity measured with respect to the systemic heliocentric velocity.
This map was made by combining all the data taken in this paper with that taken in the previous paper \citep{lim08}.  Contour levels are plotted at 3, 6, 9, 12, 15, 20, 30, 40, 50, 70, and $90 \times 0.3 {\rm \ Jy \ km \ s^{-1}}$.  
The dotted ellipse indicates where the noise has increased by a factor of 2 over that at the center.  No genuine features were detected beyond the region shown (see text).  This map was made by Hanning smoothing the corresponding channel maps by $100 {\rm \ km \ s^{-1}}$, and then discarding pixels in the channel maps with intensities less than 37~mJy.  The synthesized beam is plotted at the lower left corner.{\label{mom-all}}}
\end{center}
\end{figure}

\begin{figure}
\begin{center}
\includegraphics[width=160mm]{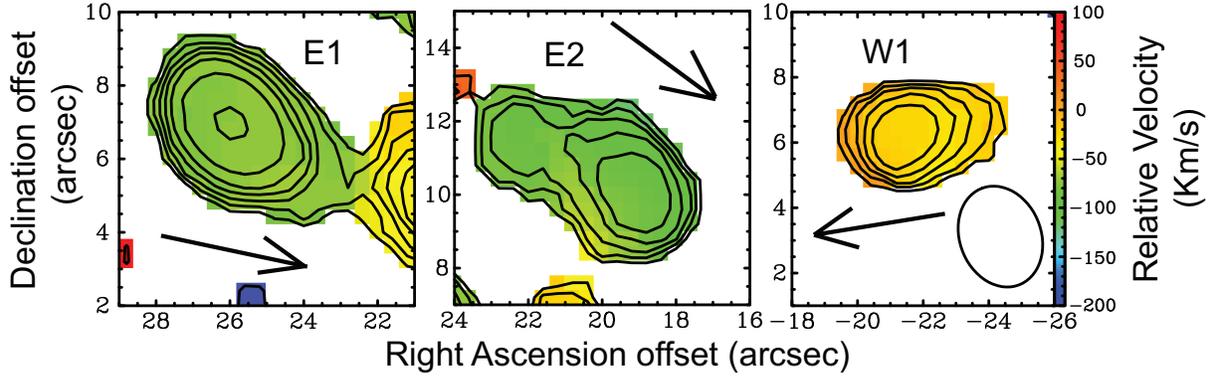}
\caption{Contours of the integrated CO(2--1) intensity and a color map of the intensity-weighted mean velocity measured with respect to the systemic heliocentric velocity for the filaments E1 (left panel), E2 (middle panel), and W1 (right panel) made from channel maps with uniform weighting.  Contour levels are plotted at 3, 6, 9, 12, 15, 20, and $30 \times 0.3 {\rm \ Jy \ km \ s^{-1}}$.  The arrows indicate the direction to the center of Per~A.  The synthesized beam is plotted at the lower right corner of the right panel.  All the moment maps were made by Hanning smoothing the corresponding channel maps by $140 {\rm \ km \ s^{-1}}$, and then for E1 and E2 discarding pixels in the channel maps with intensities less than 25~mJy and for W1 less than 20~mJy. \label{uniform-weighting}}
 \end{center}
\end{figure}

\begin{figure}
\begin{center}
\includegraphics[width=80mm]{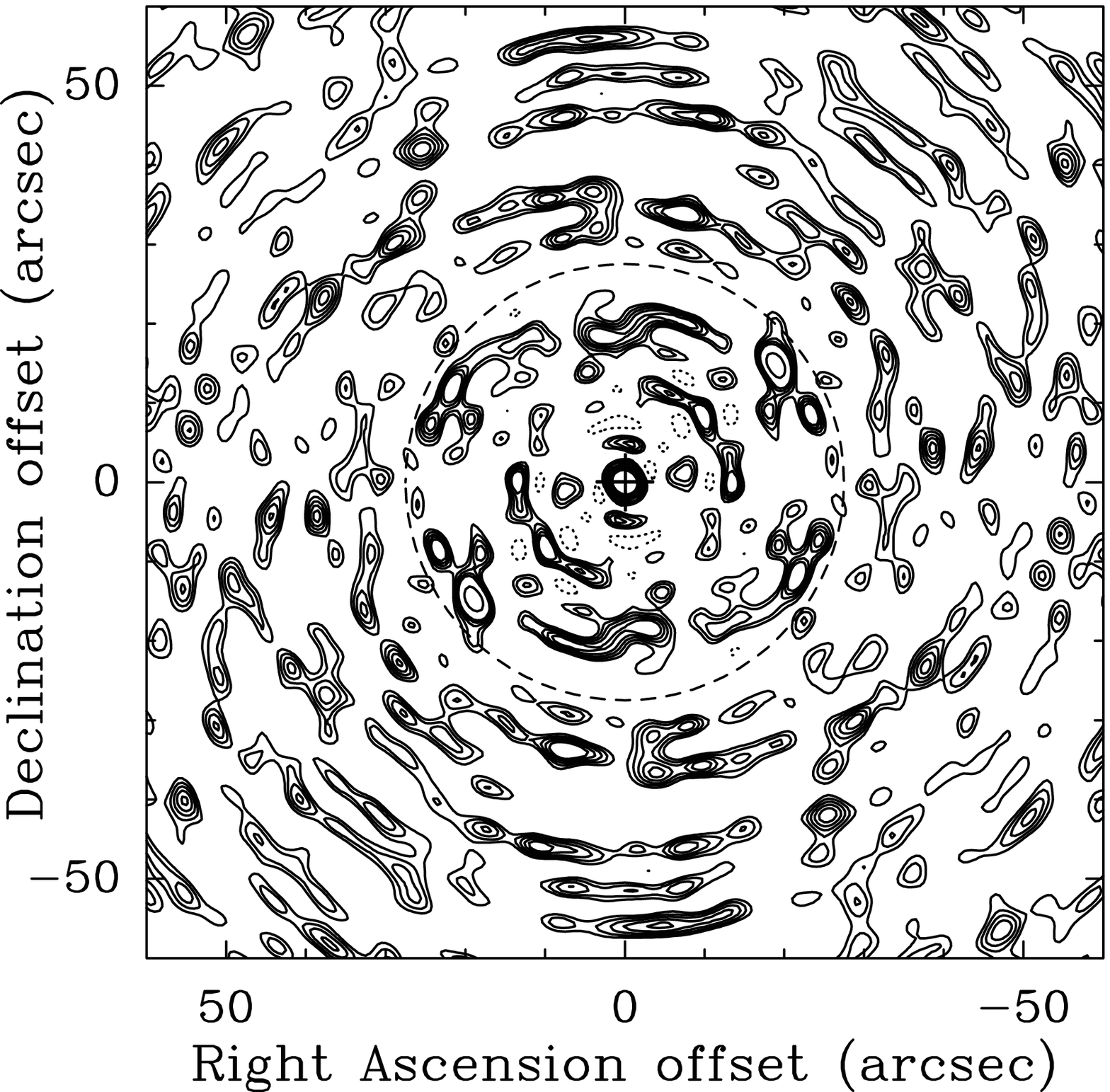}\\ 
\vspace{0.5cm}
\includegraphics[width=50mm]{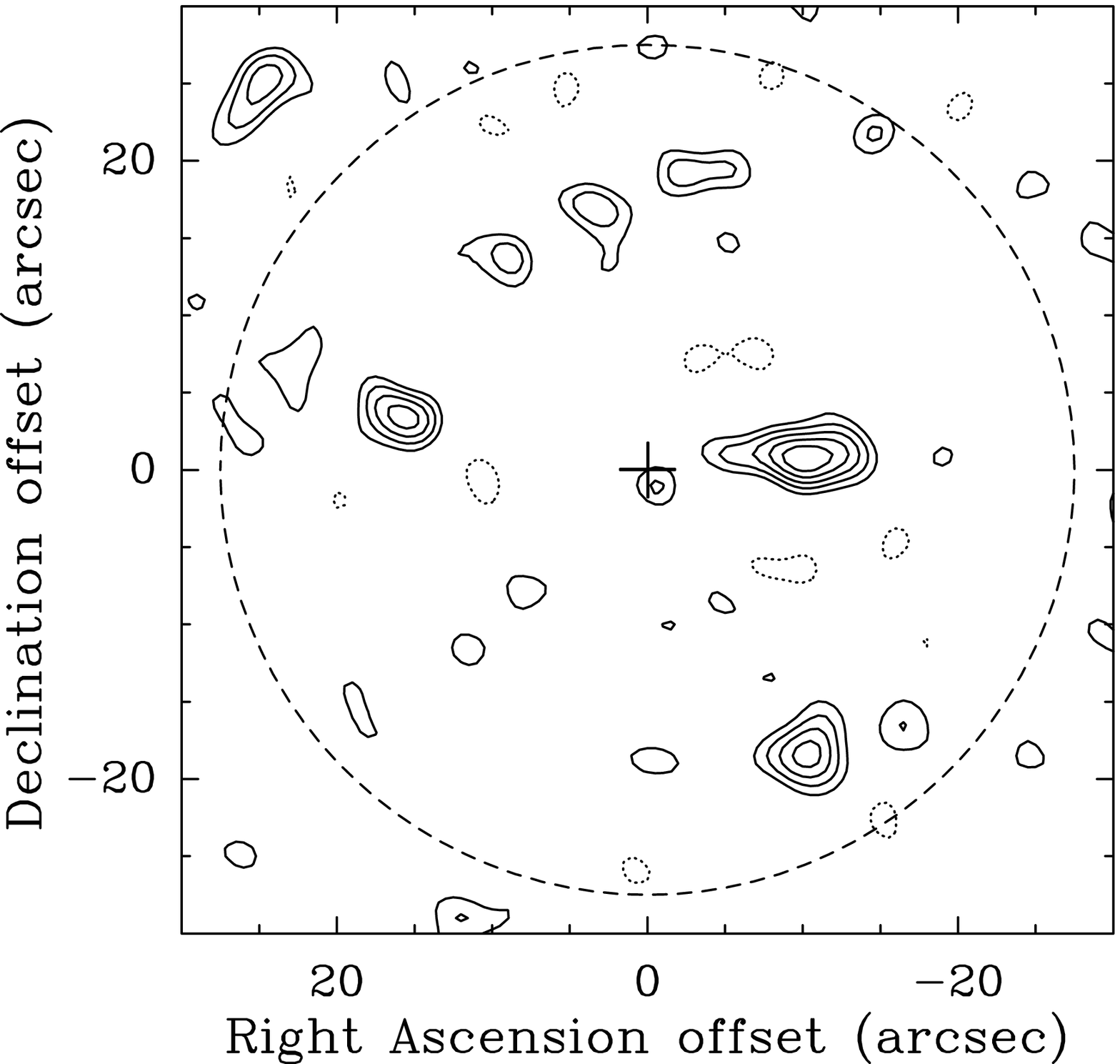}
\includegraphics[width=50mm]{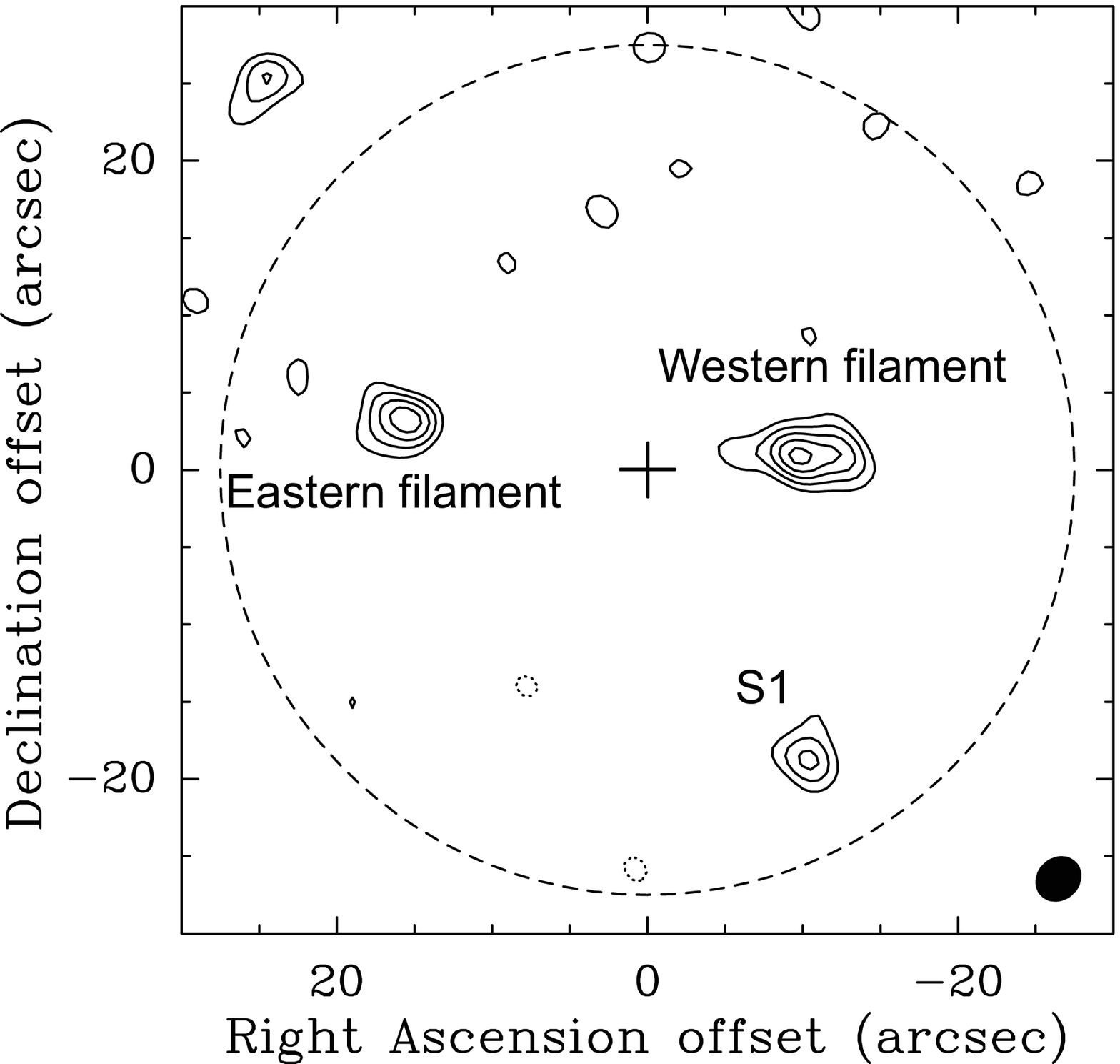}
\includegraphics[width=50mm]{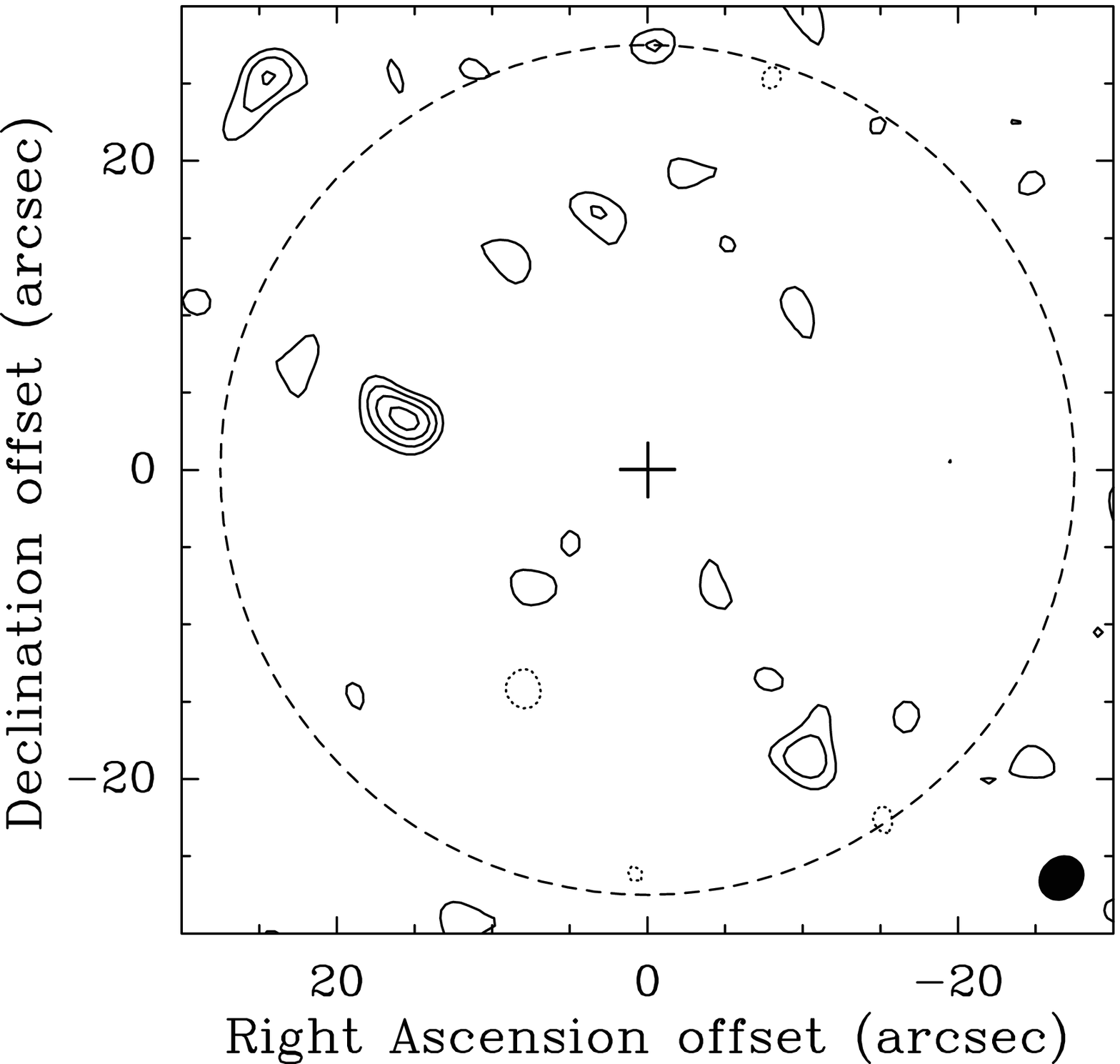}
\caption{DIRTY beam (upper panel) with contours plotted at $1\%$ $2\%$, $4\%$, $6\%$, $8\%$, $10\%$, $20\%$, $30\%$, $40\%$, and $50\%$ of the peak.  DIRTY channel map (lower left panel), CLEAN channel map (lower middle panel), and DIRTY channel map (lower right panel) after the Western filament has been subtracted from the original DIRTY map.  Contours levels in the channel maps are plotted at -3 (dashed), 2, 3, 4, 5, and $6 \times 16$~mJy. The dash circles corresponds to the FWHM of the SMA primary beam.  The synthesized beam is plotted at the lower right corner of the CLEAN map.  The channel maps are taken from Paper~I, and correspond to a velocity of $-140 {\rm \ km \ s^{-1}}$ with respect to the systemic heliocentric velocity.  The feature S1 lies at the first sidelobe of the Western filament, and remains as a significant feature after CLEANing.  After the Western filament is subtracted from the DIRTY map, however, S1 is no more intense than the low-level fluctuations, indicating that it is just a sidelobe. \label{artifact}}
\end{center}
\end{figure}

\begin{figure}
\begin{center}
\includegraphics[width=150mm]{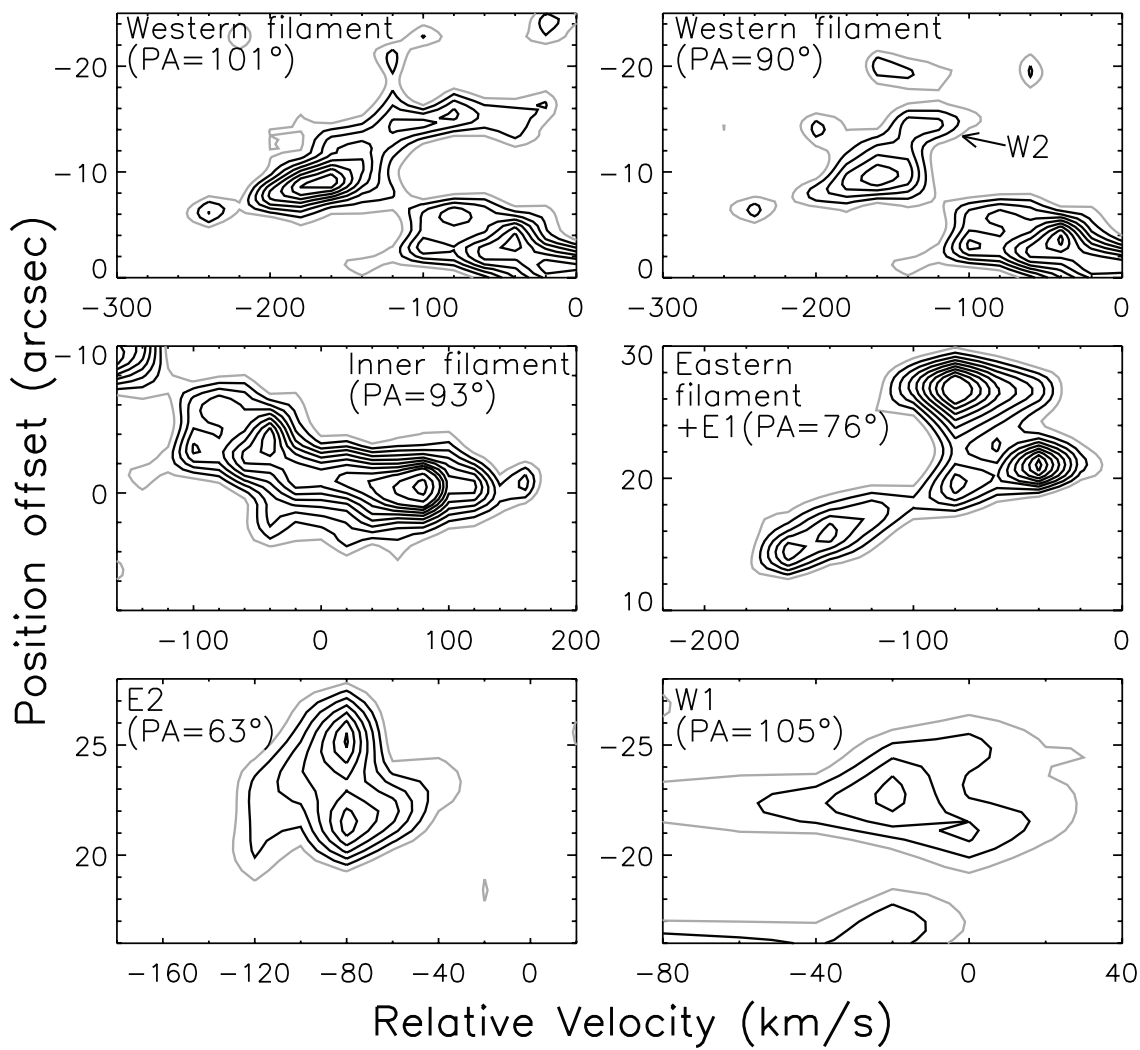}
\caption{Position-velocity (PV-) diagrams showing the CO(2-1) intensity along the radial extents of the individual filaments.  Positions correspond to offsets from the center of Per~A with negative values indicating east, and velocities are measured with respect to the systemic heliocentric velocity.
Contour levels are plotted at 2 (grey), 3, 4, 5, 6, 7, 8, 9, 10, 12 and $14 \times 15$~mJy.  \label{pv_diagram}}
\end{center}
\end{figure}

\begin{figure}
\begin{center}
\includegraphics[width=150mm]{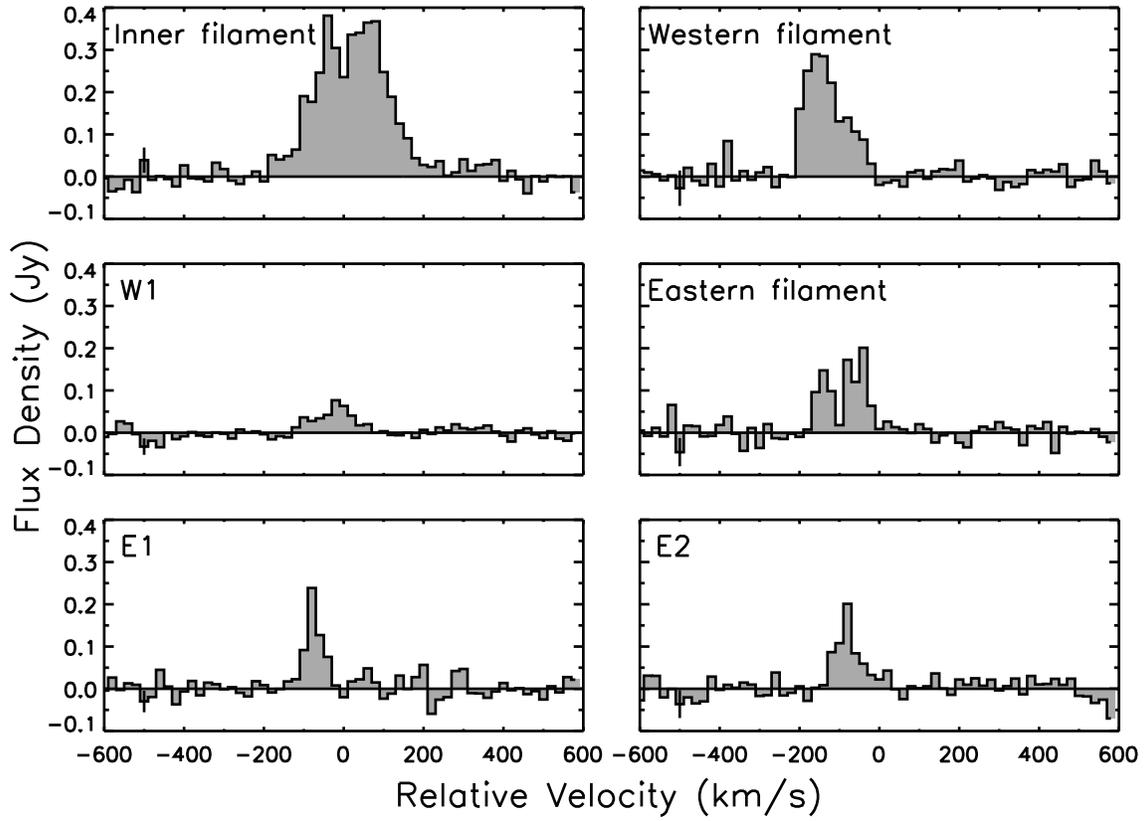}
\caption{Spectra of all the features labeled in Figure~\ref{mom-all}, derived from the channel maps shown in Figure~\ref{channel-maps}. \label{spectra}}
\end{center}
\end{figure}

\begin{figure}
\begin{center}
\includegraphics[width=90mm]{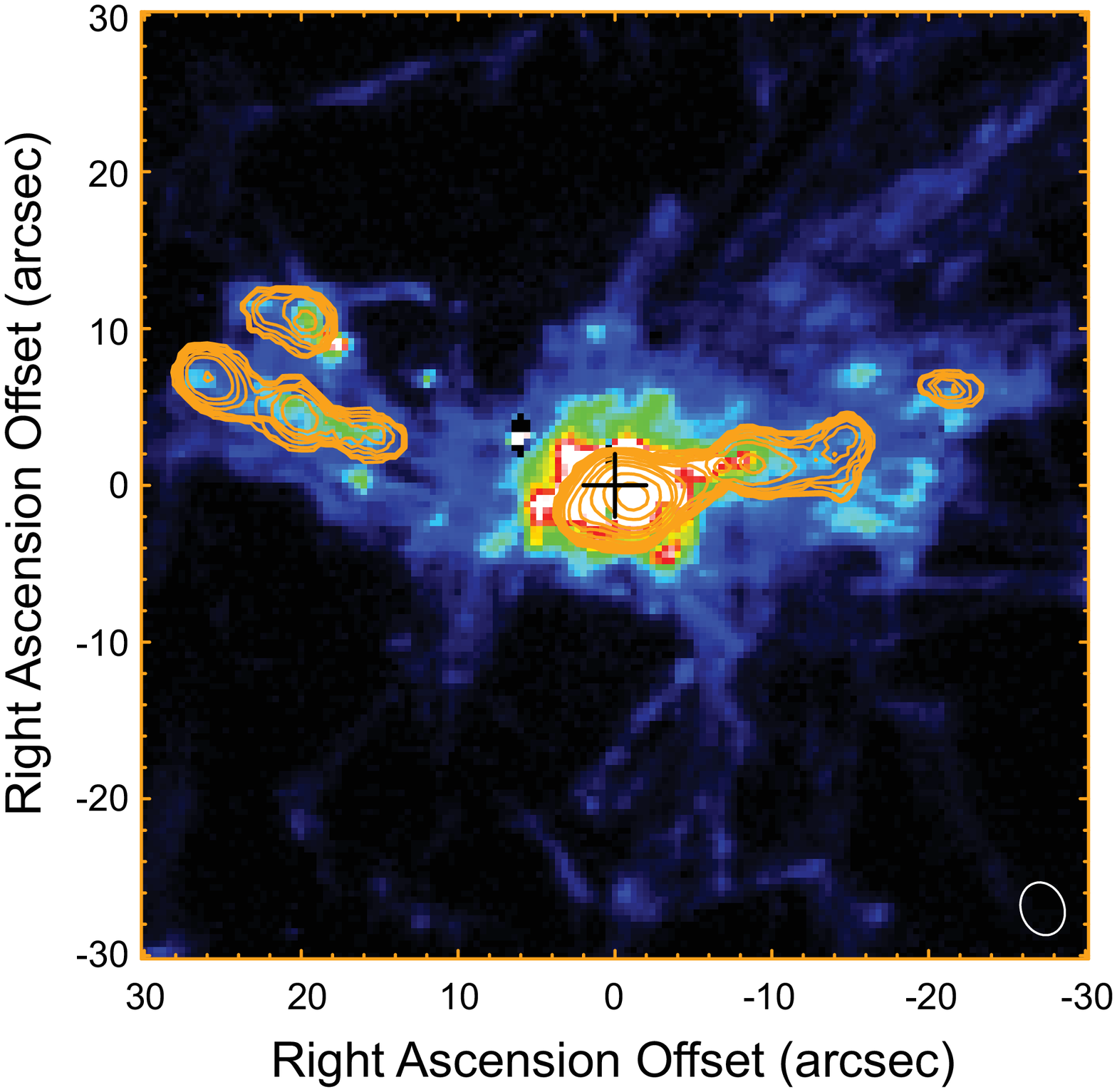}\\
\includegraphics[width=90mm]{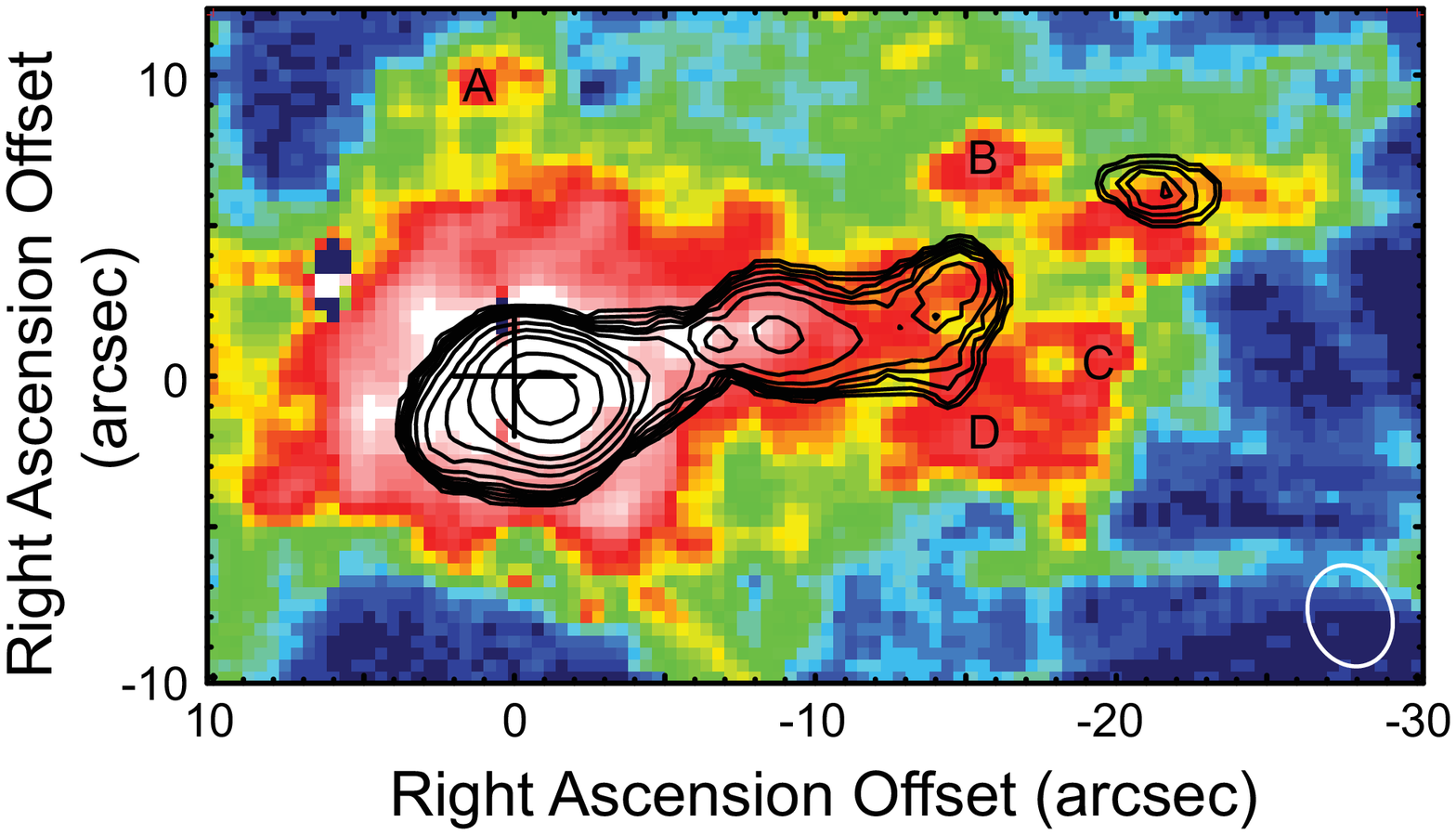}\\
\includegraphics[width=70mm]{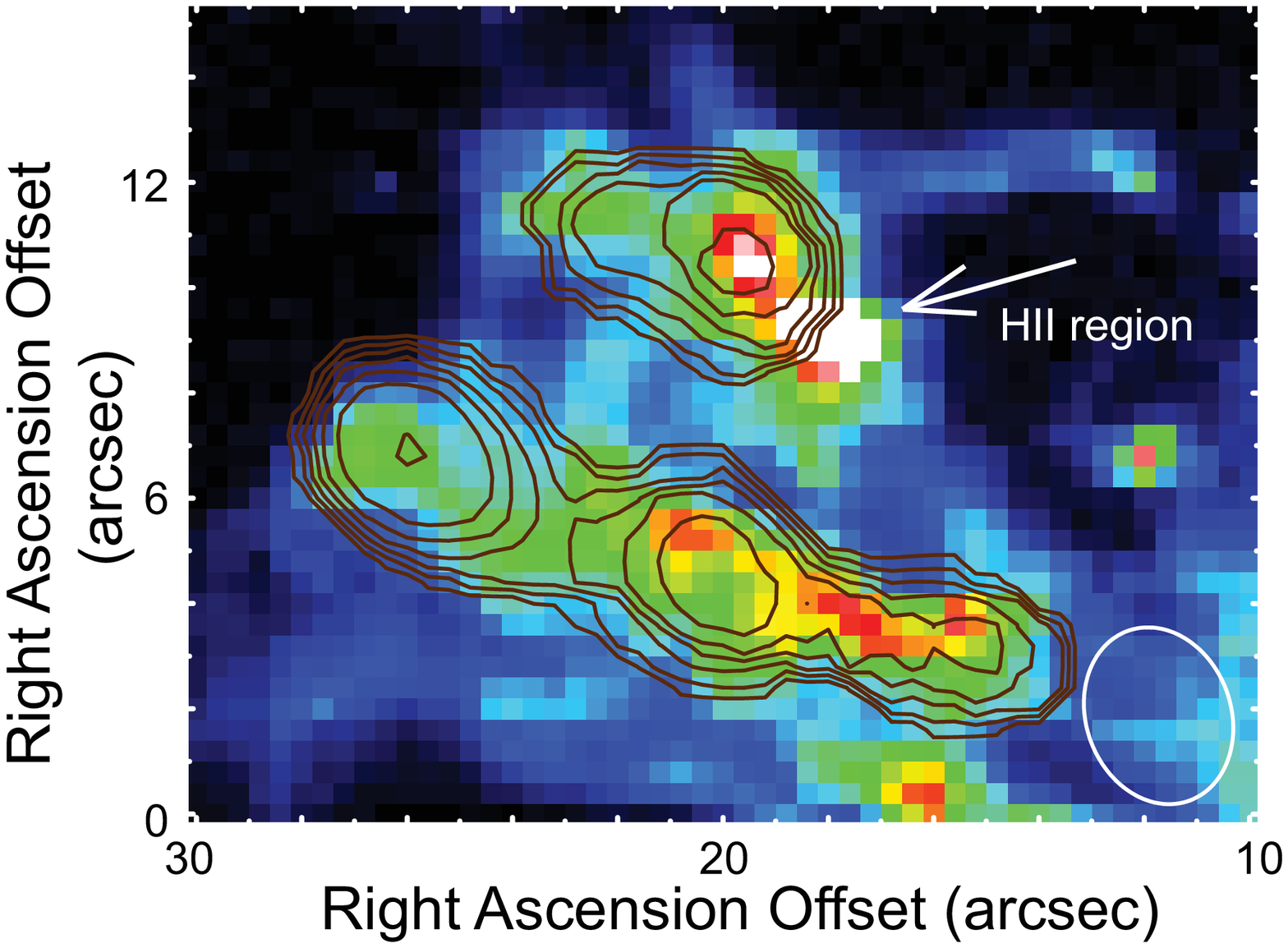}\\
{\renewcommand{\baselinestretch}{0.9}
\caption{Contours of the integrated CO(2--1) intensity (from Fig.~\ref{mom-all}, with the same contour levels plotted) superposed on a color map of the H$\alpha$$+$N[II] intensity \citep[from][]{con01}.  The top panel shows the entire region where CO(2--1) was detected.  The middle panel shows the inner and western regions, and the bottom panel the eastern region.  The H$\alpha$$+$N[II] maps are plotted with different stretches in the individual panels to emphasize different features of interest.  The synthesized beam is plotted at the lower right corner of each panel.  The H$\alpha$$+$N[II] features labeled A--D (see text) in the middle panel corresponds to those refered to in Figure~\ref{mass-haf-corr}.  The H$\alpha$$+$N[II] feature labeled HII~region in the bottom panel is that discovered by \cite{shi90}.  \label{co-haf}}}
\end{center}
\end{figure}

\begin{figure}
\begin{center}
\includegraphics[width=140mm]{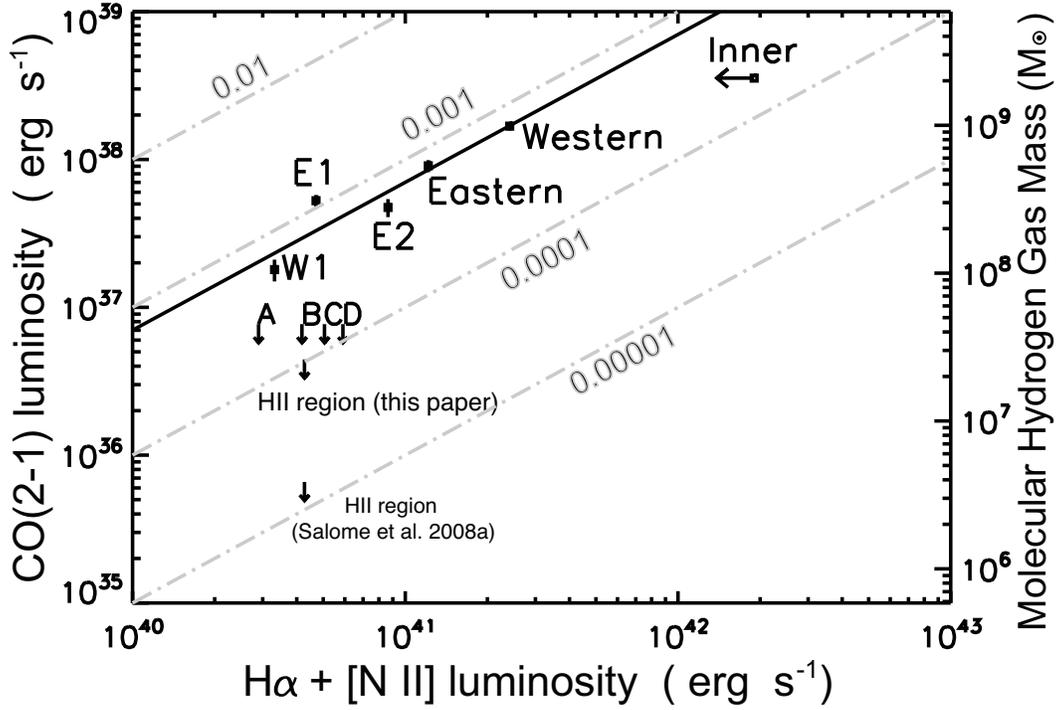}
\caption{Plot of measured CO(2--1) luminosity and derived molecular hydrogen gas mass (Table~\ref{gas_mass}) versus the H$\alpha$$+$N[II] luminosity of individual filaments.  The solid line corresponds to the best fit linear ratio of 0.0007, and the dash-dot lines to other ratios as indicated.  The features A--D and HII~region correspond to those labeled in Figure~\ref{co-haf}.  \label{mass-haf-corr}}
\end{center}
\end{figure}

\begin{figure}
\begin{center}
\includegraphics[width=90mm]{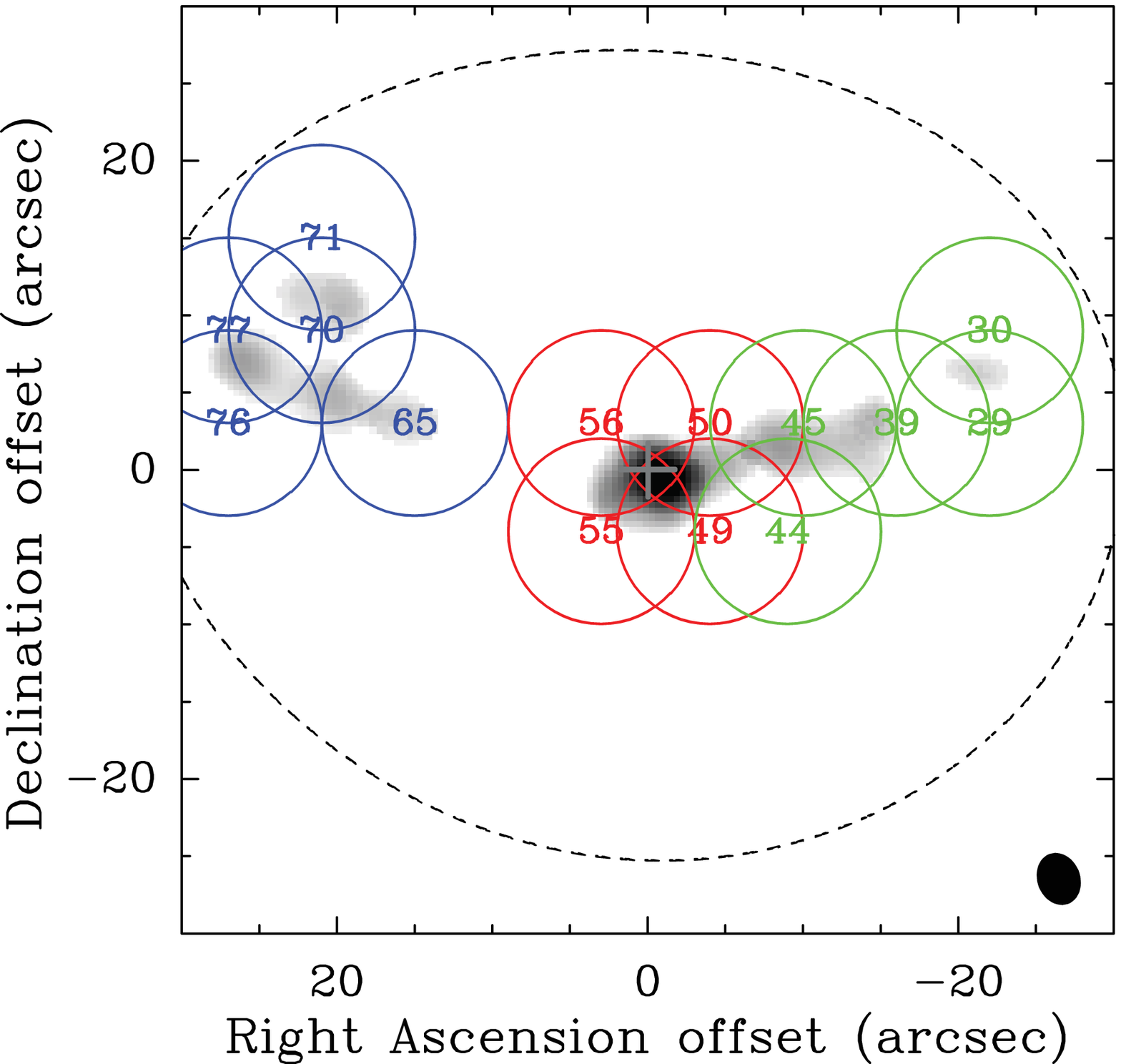}
\ \ \ \ \
\includegraphics[width=60mm]{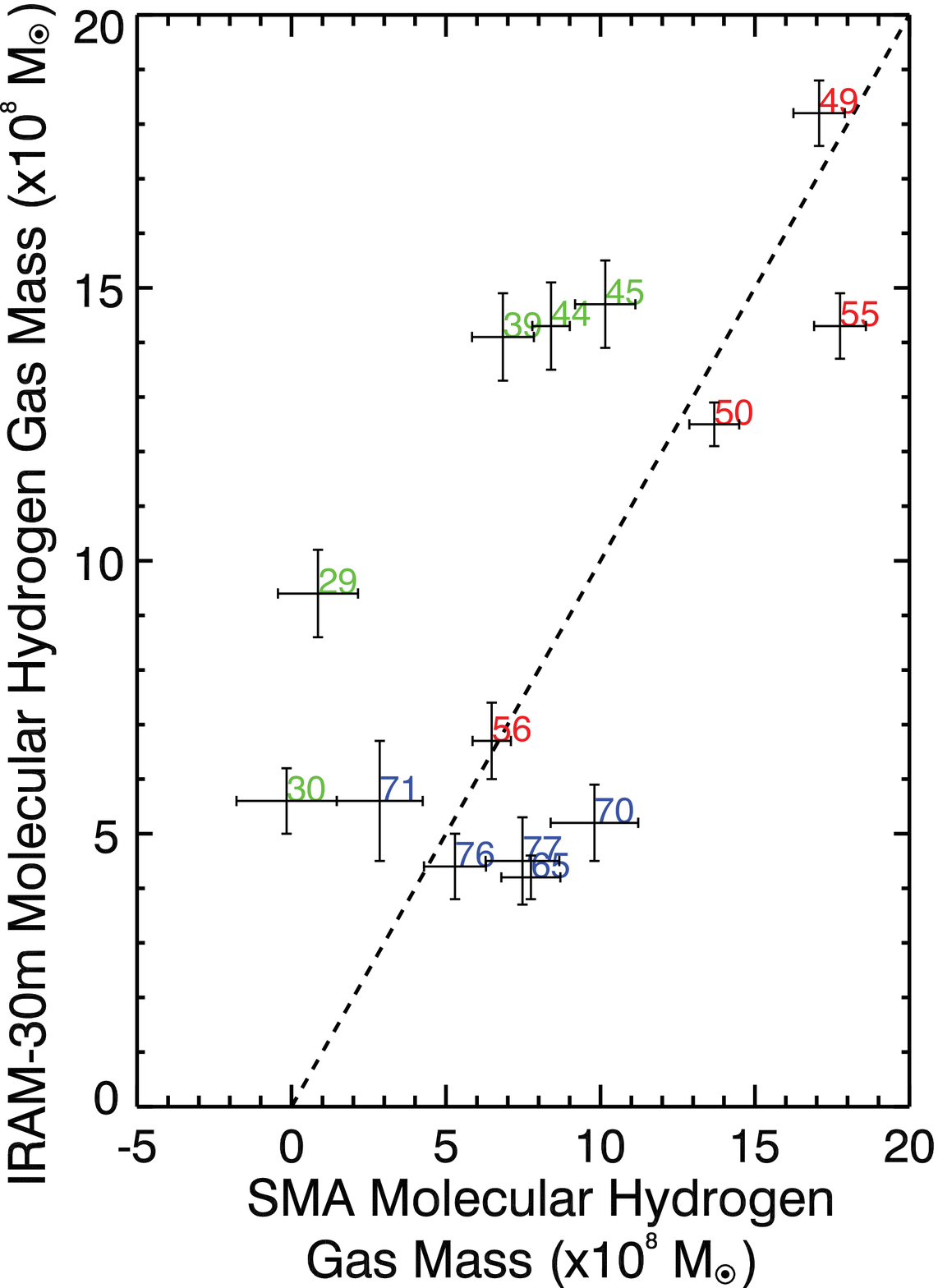}
\caption{Left panel shows the integrated CO(2--1) intensity in grayscale made from our combined data as shown in Figure~\ref{mom-all}.  The dotted ellipse indicates where the noise has increased by a factor of 2 over that at the center.  The circles correspond to the primary beams of individual IRAM-30~m telescope pointings of \citet{sal06}, with those in blue corresponding to the eastern region, those in red the inner region, and those in green the western region.  Right panel shows a plot of the molecular hydrogen gas mass measured in the individual pointings of the IRAM-30~m telescope versus that measured by us with the SMA in the same area.  The dash line indicates the same molecular hydrogen gas masses measured by the two telescopes.  Virtually all the CO(2--1) emission in the eastern and inner regions are recovered in our SMA observation, but only a fraction of that in the western regions. \label{salome}}
\end{center}
\end{figure}

We are deeply indebted to the referee for carefully reading our manuscript, picking up an important error, asking us to consider the effects of tidal shear, and suggesting other general improvements to the manuscript.  We thank Leo Blitz for bringing to our attention the nearly constant star formation efficiency found for molecular gas in nearby normal spiral galaxies, and encouraging us to compute the resultant H$\alpha$ luminosity for the molecular filaments we detect for the same star formation efficiency.  We thank Yi-Ping Ao for help in preparing the observing script, and Philippe Salom\'e for kindly providing the IRAM~30-m CO(2--1) map of Per~A that we reproduced in Figure~\ref{3pointing}.  J. Lim acknowledges a grant from the National Science Council of Taiwan in support of this work.  This grant also provides a Masters student stipend for I-.T. Ho.

{\it Facilities:} \facility{SMA}.

\newpage
\begin{table}
\begin{center}
\caption{Parameters of the Observations \label{pointing}}
\begin{tabular}{ccccccccc}
\hline\hline
Pointing&Integration time&RA&Dec&Beam size&PA&Noise\\
&(hours)&(J2000)&(J2000)&($''\times''$)&($^\circ$)&(mJy)\\
\hline
Central&$\sim$1.6&3:19:48.16&+41:30:42.10&$3.43\times2.75$&19.9&30\tablenotemark{a}\\
Eastern&$\sim$2.6&3:19:49.94&+41:30:48.10&$3.44\times2.69$&16.2&25\tablenotemark{a}\\
Western&$\sim$2.5&3:19:46.82&+41:30:44.10&$3.44\times2.66$&14.3&25\tablenotemark{a}\\
\hline
Combined&\multicolumn{3}{c}{(all three pointings and Paper~I)}&$3.43\times2.75$&19.9&15\tablenotemark{a}\\
Combined&\multicolumn{3}{c}{(all three pointings and Paper~I)}&$3.17\times2.48$&20.8&17\tablenotemark{b}\\
\hline
\hline
\end{tabular}
\tablecomments{Noise levels listed are those measured at the center of the maps.}
\tablenotetext{a}{Natural weighting.}
\tablenotetext{b}{Uniform weighting.}
\end{center}
\end{table}
\clearpage

\begin{table}
\begin{center}
\caption{Parameters of the Molecular Filaments\label{gas_mass}}
\begin{tabular}{ccccccc}
\hline\hline
Filament&Radial distance&Length&Position angle&Velocity Range&Flux Density&Gas Mass\\
&(arcsec)&(arcsec)&($^\circ$)&($\rm km \ s^{-1}$)&($\rm Jy \ km \ s^{-1}$) & ($\times 10^{8}$~M$_\sun$)\\
\hline
Inner&$\sim$3.0&$\sim$7.3&$\sim$281&-100 $\sim$ 160&$71.9\pm2.5$&$20.8\pm0.7$\\
Western&$\sim$11.8&$\sim$7.7&$\sim$281&-200 $\sim$ -40&$34.1\pm2.3$&$9.9\pm0.7$\\
Eastern&$\sim$17.8&$\sim$6.3&$\sim$76&-160 $\sim$ -40&$18.3\pm1.7$&$5.3\pm0.5$\\
W1&$\sim$22.1&$3.7\pm1.4$&$91.3\pm14.5$&-40 $\sim$ 0&$3.6\pm0.6$&$1.1\pm0.2$\\
E1&$\sim$26.5&$2.9\pm0.7$&$50.3\pm16.0$&-100 $\sim$ -40&$10.7\pm1.0$&$3.1\pm0.3$\\
E2&$\sim$23.5&$\sim$4.0&$\sim$63&-120 $\sim$ -60&$9.6\pm1.3$&$2.8\pm0.4$\\
\hline
W2&$\sim$14.5&--&$\sim$90&--&--&--\\
\hline
Total&&&&&&$43.0\pm1.2$\\
\hline\hline
\end{tabular}
\tablecomments{{Lengths of Inner, Western, Eastern, and E2 filaments derived from channel maps (see text).  Lengths of E1 and W1 are the FWHM of gaussian fits to these filaments in the uniformly weighted moment maps, and their position angles derived from these fits (see text).  Position angle of the spatially-unresolved feature W2 is its orientation relative to the nucleus.  Note that all the parameters listed have been corrected for an error in continuum subtraction and an incorrect conversion factor in Table~1 of Paper~I.  The quoted uncertainties in flux densities do not include any systematic uncertainty in absolute flux calibration that may be as large as $\sim$20$\%$.  The conversion factor used for relating CO(2--1) flux densities to molecular hydrogen gas masses is that believed to be appropriate for Galactic molecular clouds \citep{sol87}.}}
\end{center}
\end{table}

\newpage

\clearpage

\newpage
\clearpage

\clearpage

\newpage
\clearpage

\newpage
\clearpage

\newpage
\clearpage

\newpage
\clearpage

\newpage
\clearpage

\newpage
\clearpage

\end{document}